\documentclass[prb,twocolumn,english,superscriptaddress]{revtex4}
\usepackage[T1]{fontenc}
\usepackage[latin9]{inputenc}
\usepackage{verbatim}
\usepackage{float}
\usepackage{amsmath}
\usepackage{graphicx}
\usepackage{graphics}
\usepackage{times}
\usepackage[colorlinks,bookmarks=false,citecolor=blue,linkcolor=red,urlcolor=blue]{hyperref}
\usepackage{array}
\usepackage{longtable}
\usepackage{multirow}
\makeatletter

\providecommand{\tabularnewline}{\\}

\@ifundefined{textcolor}{}
{%
 \definecolor{BLACK}{gray}{0}
 \definecolor{WHITE}{gray}{1}
 \definecolor{RED}{rgb}{1,0,0}
 \definecolor{GREEN}{rgb}{0,1,0}
 \definecolor{BLUE}{rgb}{0,0,1}
 \definecolor{CYAN}{cmyk}{1,0,0,0}
 \definecolor{MAGENTA}{cmyk}{0,1,0,0}
 \definecolor{YELLOW}{cmyk}{0,0,1,0}
 }
\numberwithin{equation}{section}
\newcommand*{\rom}[1]{\expandafter\@slowromancap\romannumeral #1@}
\makeatother

\renewcommand{\vec}[1]{\mathbf{#1}}
\usepackage{babel}
\begin{document}
\title{Heisenberg antiferromagnet on Cayley trees: low-energy spectrum and even/odd site imbalance}
\author{Hitesh J. Changlani}
\affiliation{Laboratory of Atomic And Solid State Physics, Cornell University, Ithaca, NY 14853, USA}
\author{Shivam Ghosh}
\affiliation{Laboratory of Atomic And Solid State Physics, Cornell University, Ithaca, NY 14853, USA}
\author{Christopher L. Henley}
\affiliation{Laboratory of Atomic And Solid State Physics, Cornell University, Ithaca, NY 14853, USA}
\author{Andreas M. L\"auchli}
\affiliation{Institut f\"ur Theoretische Physik, Universit\"at Innsbruck, A-6020 Innsbruck, Austria}

\renewcommand{\theequation}{\arabic{section}.\arabic{equation}}

\date{March 9 2013}
\begin{abstract}
To understand the role of local sublattice imbalance in low energy spectra of $s=\frac{1}{2}$ quantum antiferromagnets, we study the $s=\frac{1}{2}$ quantum nearest 
neighbor Heisenberg antiferromagnet on the coordination 3 Cayley tree. 
We perform many-body calculations using an implementation of the 
density matrix renormalization group (DMRG) technique for generic tree graphs. 
We discover that the bond-centered Cayley tree has a quasi-degenerate set
of low lying tower of states and an {}``anomalous'' singlet-triplet finite size gap scaling. 
For understanding the construction of the first excited state from the many-body ground state, we consider a wavefunction
ansatz given by the single mode approximation (SMA), which yields a high overlap with the DMRG wavefunction. 
Observing the ground state entanglement spectrum leads us to a picture of the 
low energy degrees of freedom being "giant spins" arising out of sublattice imbalance, which helps us analytically understand the scaling of the finite size spin gap. The Schwinger Boson mean field theory has been generalized to \emph{non uniform} lattices
and ground states have been found which are spatially inhomogeneous in the mean field parameters.  
\end{abstract}
\maketitle

\section{Introduction}
\label{sec:Introduction}
Quantum antiferromagnetism for unfrustrated systems has been one of the most extensively researched subjects in condensed matter
physics. One of the simplest models in this family, the nearest neighbor
spin $\frac{1}{2}$ Heisenberg antiferromagnet on the square lattice,
	has been studied extensively: analytically with spin wave~\cite{Takahashi_SW,Hirsch_Tang_SW,Mano}
and Schwinger Boson approaches~\cite{Chubukov_SBMFT,Chubukov_Sachdev,SS_Ma,AA_PRL,AAarxiv}
and numerically with Quantum Monte Carlo~\cite{Trivedi_QMC}~(which has no {}``sign problem''
for bipartite lattices). That said, effects from physical imperfections
such as the presence of open edges~\cite{Hoglund,Metlitski} and 
static non magnetic impurities~\cite{Chernyshev_diluted_AFM} are less well understood and hence are areas of active research. 

In this paper, we study quantum antiferromagnetism on the Cayley tree~(or Bethe Lattice~\cite{FN_Bethe}) \mbox{--} a bipartite lattice without loops, with the motivation of understanding the low energy spectrum of the spin half Heisenberg model on this lattice. 
The Cayley tree has the well-known general pathology that, in the thermodynamic limit, the number of 
boundary sites is a finite fraction of the total number of sites;
as a consequence, different ways of approaching the
thermodynamic limit may give different results
in any problem based on the Cayley tree.
A particular manifestation of this for the antiferromagnet is
that finite trees may have a large
excess of sites belonging to one sublattice over the other.

Our systematic study of the low energy spectrum of the spin half Heisenberg Hamiltonian on the Cayley tree shows that the effect of sublattice imbalance is to create a "tower of states", lower than the Anderson tower of states~\cite{Anderson,Lhuillier,Sachdev,Ziman,Gross}. A similar result was obtained by Wang and Sandvik~\cite{Wang_Sandvik_1} from their study of spin half antiferromagnets on diluted square lattices.
 Aided by numerical calculations, we propose a framework for understanding this effect. We also find that Schwinger Boson Mean Field theory~\cite{AA_PRL} is
a good description of the many body ground state and can reproduce many of its features quantitatively.

Previous studies of this model by Otsuka~\cite{Otsuka} and
Friedman~\cite{Friedman} focused primarily on ground state properties and excited states were not considered
in these studies. More recently Kumar et al.~\cite{Kumar_Ramasesha_Soos} 
have significantly extended this analysis to both the spin-1/2 and spin-1 Heisenberg model. We use all these studies as useful benchmarks for our own 
numerical calculations. 

From a theorist's perspective, the Cayley tree achieves 
many simplifications which makes exact solutions possible, e.g.
the Bose Hubbard model
on this lattice was recently solved by Semerijan, Tarzia and Zamponi~\cite{Semerijan}.
It is also the basis of approximations such as
the Brinkman-Rice treatment of the Hubbard model~\cite{Brinkman_Rice}.
More recently, it found applications in the treatment of the quantum impurity
problem which is at the heart of dynamical mean field theory (DMFT)~\cite{DMFT}. 
It does not appear that a spin model has been 
realized on such a topology experimentally (though there has been interest in the study of dendrimers~\cite{Laguna,Dendrimer_1,Dendrimer_2}).

In our case, the complete absence of loops makes this lattice conducive for 
the density matrix renormalization group (DMRG) algorithm~\cite{White}.   
With the DMRG method we have an explicit (yet compact) representation
of ground and excited state many-body wavefunctions which gives us
abundant information to understand the low energy properties of these
systems. In particular, reduced density matrices can be used as tools
to understand properties of these states~\cite{Muender}.  

The remainder of the paper is divided as follows. In section~\ref{sec:Model} we 
introduce the model and lattices being investigated and define a measure of sublattice imbalance associated with them. 
In section~\ref{sec:DMRG}, we give a brief overview of our implementation of the DMRG algorithm
applied to generic trees. In section~\ref{sec:GS}, we discuss the general properties
of the ground state and excited states of the bond-centered Cayley
tree. In section~\ref{sec:SMA}, we use a variational ansatz given by the single
mode approximation, in conjunction with an argument from first order perturbation theory, to explain the finite size scaling of the spin
gap. Finally, in section~\ref{sec:SBMFT}, we corroborate our observations of the ground
state properties, by the use of the Schwinger Boson mean field theory (SBMFT). 

\section {The Model}
\label{sec:Model}
We consider the nearest neighbor antiferromagnetic spin 1/2 Heisenberg model 
with uniform coupling $J$, 
\begin{equation}
H=J\sum_{\left\langle i,j \right\rangle}\vec S_{i} \cdot \vec S_{j} \label{eq:Heisenberg_Hamiltonian}
\end{equation}
In this paper, we use the spin rotational symmetry of the Hamiltonian (\ref{eq:Heisenberg_Hamiltonian}),
to label many body states by $\left| S,S_{z} \right\rangle$, where $S$ refers to the spin of the state and $S_{z}$ is its $z$ component. 

On a bipartite lattice (with sublattices \emph{A} and \emph{B}), like the Cayley tree, 
with $n_{A}$ sites in sublattice \emph{A}
and $n_{B}$ sites in sublattice \emph{B}, it is rigorously known~\cite{Hulten_Marshall_Lieb_Mattis}
that the ground state of the Heisenberg Hamiltonian has a net spin 
$S=\left|\frac{n_{A}-n_{B}}{2}\right|$. 

\begin{figure}[htpb]
\centering
\includegraphics[width=0.8\linewidth]{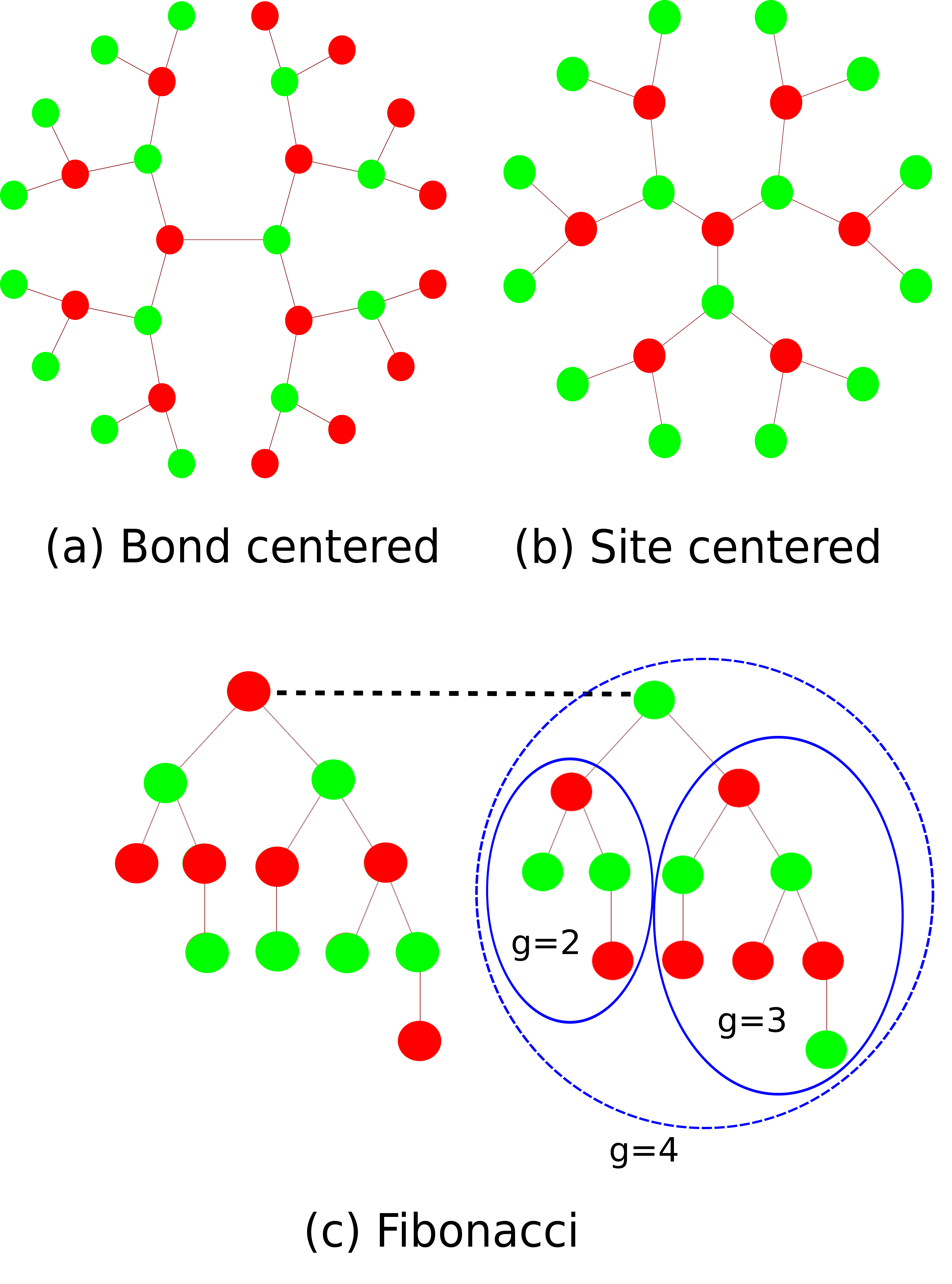}
\caption{(Color online) (a) The bond-centered Cayley tree. 
(b) The site-centered Cayley tree. 
In both cases all sites, other than those on the boundary, have coordination 3.
(c) The "Fibonacci Cayley tree" is constructed hierarchically and has some coordination 2 sites. 
The figure shows a generation 4 cluster constructed by connecting the roots (head sites) 
of the generation 2 and generation 3 trees to a common site (the root of the generation 4 tree). 
To have a globally balanced cluster we introduced a bond connecting the root of the generation 4 tree with the root of its mirror image. 
All clusters in (a),(b),(c) are bipartite (the dark~(red) and light~(green) colors show the two sublattices) 
and have no loops. 
}
\label{fig:bond_site_center}
\end{figure}

The first kind of tree we consider is the {}"bond-centered" Cayley tree of the form depicted
in Fig.~\ref{fig:bond_site_center}(a). 
The number of sites $N_{s}$ for such a cluster is related 
to the {}``generation'' $g$ by,
\begin{equation}
N_{s}(g)=2^{g+1}-2\label{eq:N_S_bond_centered}
\end{equation}
Since the bond centered clusters have no {}"global imbalance" i.e. $n_{A}=n_{B}$, 
the ground state is a singlet (and the monotonicity of the energy with total spin $S$ 
implies that the first excited state is a triplet). 

As mentioned before, the notion of {}"local sublattice imbalance" will be crucial 
in understanding the low energy physics. 
For the bond centered cluster, we define a measure of imbalance (which we refer to as $I_{b}$ from here on) 
by dividing the cluster at the central bond into two equal parts. 
We count the excess of one sublattice over the other in one half of the cluster and multiply by 1/2 for spin 1/2. 
It can be easily shown that $I_{b}(g)$ is related to the generation $g$ as,
\begin{equation}
I_{b}(g)=\frac{2^{g}\pm1}{6}\label{eq:Imbalance_bond_centered}
\end{equation}
where $+(-)$ is for $g$ odd(even).

Fig.~\ref{fig:bond_site_center}(b) is the more usual way of defining a Cayley tree and which
we refer to as {}``site-centered''. The number of sites is related to the generation $g$ by,
\begin{equation}
N_{s}(g)=3\left(2^{g}-1\right)+1\label{eq:N_S_site_centered}
\end{equation}
Unlike the bond centered cluster, a global sublattice imbalance exists here 
which leads to a ground state spin of $S_{0}\equiv2^{g-1}$. 
We measure the imbalance $I_{s}(g)$ 
in either of the three symmetrical arms of the site centered Cayley tree: specifically, we 
count the excess sites of one sublattice over the other (in one arm) and multiply by 1/2. 
This definition is particularly convenient as it gives us $I_{s}(g)=I_{b}(g)$ for all $g$.

A recent publication on the study of the Heisenberg
model on the Cayley tree by Kumar, Ramasesha and Soos~\cite{Kumar_Ramasesha_Soos}
considers the site-centered clusters. We confirm their results 
for the site-centered case, but interestingly find that the bond-centered cluster
has significantly different ground and excited state properties. 
We will provide some brief comparisons in section~\ref{sec:GS} to illustrate this point.

How is the situation different if there is no imbalance locally? To address this, we introduce the 
{}"Fibonacci Cayley tree". The recipe for constructing the generation
$g+1$ Fibonacci-Cayley tree is to combine the generation $g$ and $g-1$ trees by connecting their roots (head sites) to a common site (which in turn serves as the root (head site) of the generation $g+1$ tree). 
Fig.~\ref{fig:bond_site_center}(c) illustrates this construction.
 
If we label the number of odd and even sublattice sites by $A_{g}$ and $B_{g}$ respectively,
then (counting the root as even), we get,
\begin{subequations}
\begin{eqnarray}
A_{g+1} & = & 1+B_{g}+B_{g-1}\label{eq:A_g}\\
B_{g+1} & = & A_{g}+A_{g-1}\label{eq:B_g}
\end{eqnarray}
\end{subequations}
The total number of sites $N_{s}$ at generation $g+1$ is,
\begin{eqnarray}
N_{s}(g+1) &=& A_{g+1}+B_{g+1}               \nonumber \\
	   &=& 1+B_{g}+B_{g-1}+A_{g}+A_{g-1} \nonumber \\
	   &=& 1+N_{s}(g)+N_{s}(g-1)         \label{eq:Fibo_rec}
\end{eqnarray}
Observe that $N_{s}(g)$ satisfies the Fibonacci recursion, that is, 
$N_{s}(g) = F_{g+1}-1$, where $F_g$ is the $g$-th Fibonacci number, which justifies the name of the tree.
The size of this lattice grows as $\tau^{g}$ where $\tau$ is the
golden ratio $\tau=(1+\sqrt{5})/2\sim1.618$. Also, every 3rd generation
is unbalanced by 1 and every other generation is both globally and
locally balanced.  
Table~\ref{table:fibonacci} shows the sizes of the Fibonacci-Cayley clusters 
along with the number of sites in the even and odd sublattices for up to $g=11$ generations.  
\begin{table}[htpb]
\begin{center}
\begin{tabular}{l | c | c | c | c | c | c | c | c | c | c | c | c}
\hline
$g$&      0& 1& 2& 3& 4& 5& 6& 7& 8& 9& 10& 11 \\
$A_g$&    1& 1& 2& 4$*$& 6& 10& 17$^*$& 27& 44& 72$^*$& 116& 188\\
$N_s(g)$& 1& 2& 4& 7&   12& 20& 33&  54&  88& 143& 232& 376 \\
\hline
\end{tabular}
\caption
{Number of sites in a Fibonacci-Cayley tree as a function of generation $g$.
$A_g$ is number of $A$ sublattice sites;
$B_g=A_g$ except for the entries marked with $*$, in which case $B_g=A_g-1$.
The total count is $N_s(g)=A_g+B_g$.}
\label{table:fibonacci}
\end{center}
\end{table}

In order to have a balanced cluster at every generation, we combine two identical generation $g$ Fibonacci constructions 
(as in equation~(\ref{eq:Fibo_rec})), by introducing a bond connecting their roots as 
shown in Fig.~\ref{fig:bond_site_center}(c). 

\section{Density Matrix Renormalization Group On Generic Tree Graphs}
\label{sec:DMRG}
The density matrix renormalization group (DMRG) is a powerful numerical
technique for studying many-body systems. It was developed by White~\cite{White} for one dimensional systems to remedy the problems associated with the Numerical Renormalization Group (NRG)~\cite{Wilson}. DMRG has also been generalized to study lattice models (such as the spin 1/2 Heisenberg model~\cite{Otsuka,Friedman,Kumar_Ramasesha_Soos} and the fermionic Hubbard~\cite{Lepetit} model) on the Cayley tree. More recently, Murg et al.~\cite{Murg_TTN} have also used the Cayley tree to embed quantum
chemical systems to study them with tree tensor network algorithms (closely related to the DMRG). 

When Otsuka~\cite{Otsuka} and Friedman~\cite{Friedman} first adapted the DMRG 
to the Cayley tree, their procedure was for regular trees and utilized an
infinite-system DMRG. A recent publication by Kumar et al.~\cite{Kumar_Ramasesha_Soos} (for the site-centered lattice) 
improves upon the scaling of previous algorithms, 
by considering a efficient way to hierarchically construct the lattice. 
Our implementation differs from all of the above, as it allows us to study 
the properties of \emph{any} finite tree (and not necessarily regular ones, eg. percolation clusters~\cite{Changlani}).
We outline the details in the remainder of this section.

We spell out the notation we have used in this section. $d$ is the number of degrees of freedom per site. 
For example, $d=2$ for a spin 1/2 Hamiltonian.
$z$ is the coordination number of a site 
(Although our discussion talks in terms of uniform $z$, a generalization to site dependent $z$ is straightforward). 
$M$ will be used to denote the number of retained states on a local cluster of sites (or {}"block").

\subsection{Initialization}
\label{sec:DMRG_Initialization}
\begin{figure}[htpb]
\centering
\includegraphics[width=\linewidth]{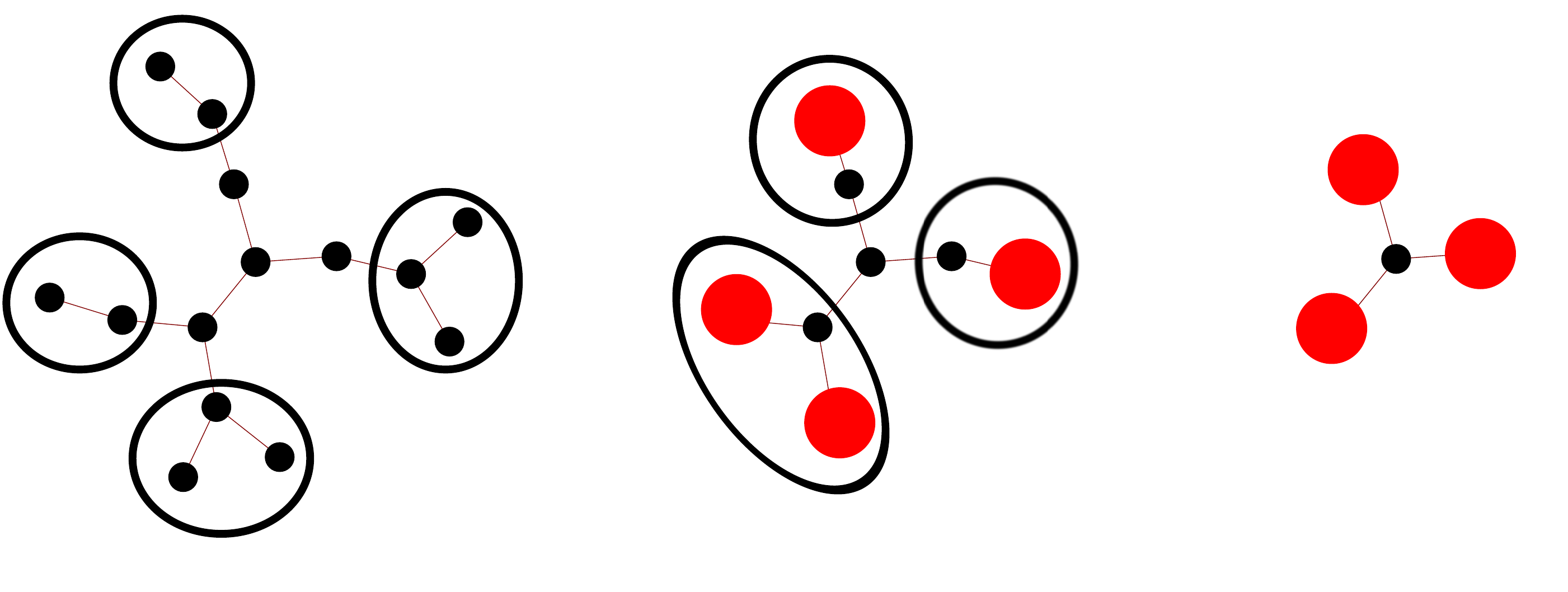}
\caption{Warm-up step of the DMRG involving multiple independent renormalization
procedures on the tree utilizing energy based truncation. The {}``army
continues to march in'' from all sides till one reaches the geometrically
central point (often called the {}``focal point" or {}``root'').
Here we show all stages for a given tree. The red dots represent renormalized blocks.}
\label{fig:nrg}
\end{figure}

Our algorithm starts with the generation of a suitable guess wavefunction.
As Fig.~\ref{fig:nrg} shows, this is done by performing "energy based truncations" of the Hilbert space
on a successive hierarchy of clusters of sites (or {}"blocks"), assuming they are completely disconnected from the rest of the cluster. 
By {}"energy based truncation" we mean that on each block we retain only the $M$ lowest energy states of the local block Hamiltonian, 
where $M$ is a parameter that determines the accuracy of the calculation. 

This blocking procedure is carried out at various parts of the tree, beginning from the boundary sites of the cluster and terminating when one reaches the geometrically central point (called the {}``focal'' point or {}``root''). 
Thus, at the end of the initialization calculation, 
one has a description of the Hamiltonian of the entire system in terms of the 
root degree of freedom surrounded by $z$ blocks. 

While the end result of our calculations are expected to be independent 
of this initialization, a good choice for the starting guess 
wavefunction can greatly accelerate convergence. 
In particular, we modify our criterion for retaining states when targeting an excited state (say in a high $S_{z}$ sector). During the initialization we introduce a {}"fake" uniform magnetic field in the Hamiltonian 
to favor retention of states that describe the high energy wavefunction.   

\subsection{Density matrix based truncation}
\label{sec:DMRG_truncation}
We will now consider how every {}"iteration" in the DMRG is carried out to systematically approach the ground (or excited) state(s) of the system.
For this purpose, we require a description of the full Hamiltonian in terms of a {}"site" degree of freedom 
(here $(\uparrow,\downarrow)$) and the basis spanned by the $M^{z}$ states 
retained on the $z$ blocks surrounding it.

At every iteration we use the Lanczos algorithm~\cite{Lanczos} to solve
for the lowest energy eigenvector of the entire system (also
referred to as the "superblock"). Then, treating one of the
blocks as the "environment" and the remaining $z - 1$ blocks
(the "systems") and the "site" collectively as the "new system",
we obtain the reduced density matrix, $\rho_{\text{RDM}}$, of the "new
system" from the ground state ($\psi_{GS}$) of the superblock by
computing,
\begin{equation}
\rho_{\text{RDM}}\equiv \rm Tr_{env}\left(\left|\psi_{GS}\left\rangle \right\langle \psi_{GS}\right|\right)\label{eq:RDM}
\end{equation}
As is illustrated in Fig.~\ref{fig:sweep}, each block takes its turn being the {}"environment"
while the other blocks together with the {}"site" act as the system.
(The order of choosing environments is not very crucial to the final result.) 

In addition to the ground state density matrix, we have also targeted higher excited states, by performing a state
averaging of the reduced density matrix,
\begin{equation}
\rho_{\text{RDM}}^{\text{avg}} \equiv \frac{\sum_{i}w_{i}\rm Tr_{env}\left(\left|\psi_{i}\left\rangle \right\langle \psi_{i}\right|\right)}{\sum_{i}w_{i}}\label{eq:state_averaged_RDM}
\end{equation}
where $w_{i}$ is the positive weight given to the density matrix
formed from state $\left|\psi_{i}\right\rangle $. In most cases,
we simply used $w_{i}=1$ for all states we targeted. 
An advantage of state averaging is that it often 
helps us to prevent (or get us out of) solutions which are local (but not global) minima. 

The reduced density matrix is diagonalized and only $M$ states
with the highest eigenvalues (out of the total $dM^{z-1}$ states) are
kept on the block. 

\subsection{Sweep algorithm}
\label{sec:DMRG_sweep}
\begin{figure}[htpb]
\centering
\includegraphics[width=0.9\linewidth]{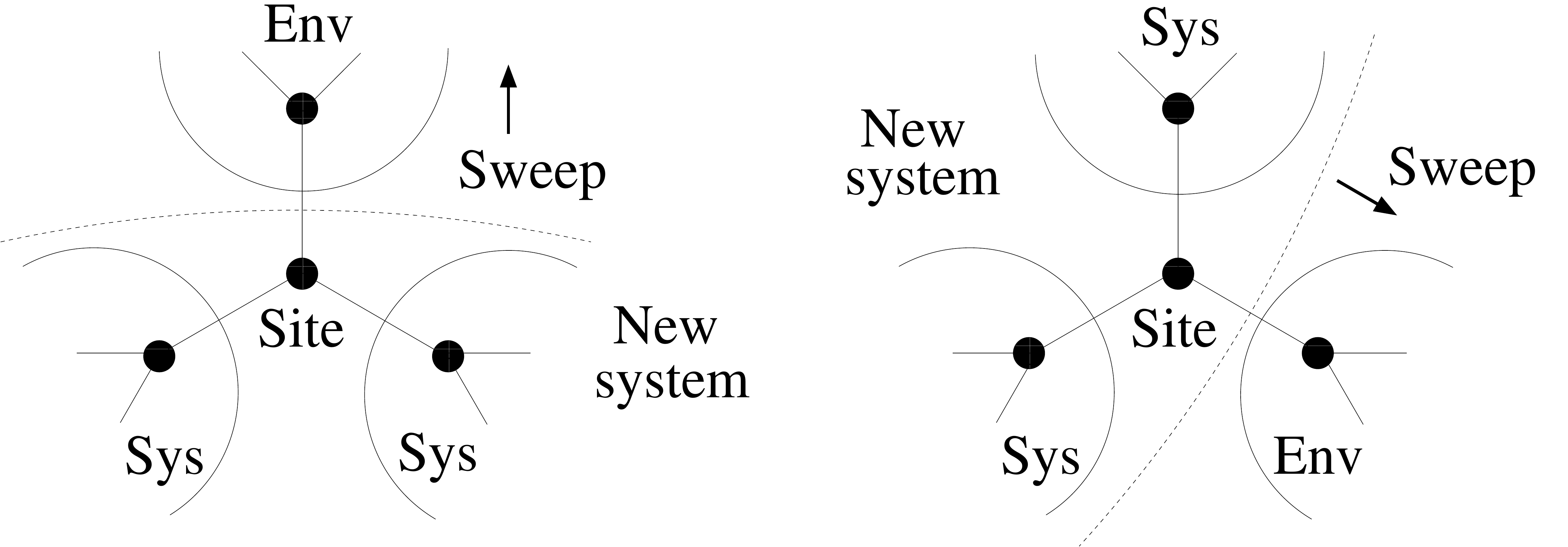}
\caption{Division of the Cayley tree locally into site, system(s) and environment as required by
the DMRG. One renormalization step consists of combining the site
and system(s) into a new system, retaining the states governed by
tracing out the environment degrees of freedom.}
\label{fig:sweep}
\end{figure}

Once the density matrix based truncations with the root as {}"site" are completed, 
the algorithm proceeds to consider a new division of the lattice 
into {}"systems" and {}"environment" by considering a site on the shell one step out from the root of the tree.
Each of the $z$ sites connected to the original root gets its turn being the "new root".(See the arrows in Fig.~\ref{fig:sweep} 
as an example of the directions in which the "sweep" algorithm proceeds).

After this stage, we consider the sites which are
two steps away from the root of the tree (i.e. one step away from the previous sites used to divide the system). This
"sweeping out" process continues till one reaches the exterior of the cluster.

Once we reach the exterior of the cluster, we {}"sweep in", like we did during the initialization procedure. However, this time (and for all future sweeps) 
we have an environment present, whose states we trace over to guide the density matrix based truncation process. This in-out-in sweep
continues till convergence of the energy is attained.  
One "complete sweep" is defined here as a "sweep out" (from the root to the boundaries) followed by a "sweep in" (from boundaries to the root) operation.

The scaling of algorithm (per sweep) can be understood as follows. Each Lanczos diagonalization of the superblock costs $M^{z}d \times Md$ 
amount of effort and there are $N_{s}$ such diagonalizations needed, 
where $N_{s}$ is the number of sites in the tree. 
For a highly symmetrical lattice (such as the Cayley tree), one can reduce this computational cost to 
$\ln(N_{s}) M^{z+1}d^{2}$ (however this has not been implemented).

The reduced density matrix computed from the eigenvector of the superblock has dimensions $M^{z-1}d \times M^{z-1}d $, which costs $M^{3(z-1)}d^3$ amount of computational effort to diagonalize. However, one must keep in mind that this reduced density matrix has a block diagonal structure owing to the fact that the retained states have definite $S_z$ symmetry, which brings down the cost associated with this step.   

\subsection{Computing Expectations}
\label{sec:DMRG_Expectations}
Just as in the usual one dimensional DMRG, we have an explicit representation
of the wavefunction in terms of the block states and we can efficiently
compute various kinds of correlation functions. For the purpose of
this paper, we have simply measured the spin-spin correlation functions
$\left\langle \vec S_{i}\cdot\vec S_{j}\right\rangle $ and the
matrix element $\left\langle 1 \right|S_{i}^{+}\left|0\right\rangle$, where $\left| 0 \right\rangle$ ($\left| 1 \right\rangle$) 
refers to the ground state singlet (first excited state triplet) and has the labels $\left|S=0,S_{z}=0\right\rangle$ ( $\left|S=1,S_{z}=1\right\rangle$).
Both these functions are needed for calculating the coefficients occurring
in the single mode approximation, which will be discussed in section~\ref{sec:SMA}. 
The latter is computed by targeting the ground and excited state in the same DMRG run, 
so that both states have the same block representation. 

We also compute the eigenvalues of the reduced density matrix (for a particular division of the lattice), collectively known as the 
{}"entanglement spectrum". These turn out to be a 
very useful probe of the low energy degrees of freedom 
(as we will see in section \ref{sec:SMA_Giant_Spin}). This needs no extra effort in the DMRG, 
since these eigenvalues are computed anyway as part of the Hilbert space 
truncation process. 

\subsection{Parameters and Benchmarks}
\label{sec:DMRG_benchmarks}
All calculations reported here are for trees with a maximum coordination of $z=3$. For all systems considered here, the retained number of states $M$ was less than or equal to 160 and we observed a reasonable convergence of the energy within 10-20 sweeps. 

To benchmark our calculations we have also compared our results with Exact Diagonalization data where possible. 
In particular, for the bond centered tree we considered the ground state energies and correlations in all $S_{z}$ sectors of the 30 site cluster
and some high $S_{z}$ sectors for the 62 site cluster.

One can see from Table~\ref{table:190_176} that the convergence of the energy (in the $S_z$ sector corresponding to the spin $S_0$ of the ground state) 
and the spin gap (defined to be $E(S_0+1)-E(S_0)$) is rapid as a 
function of the number of retained states on a block ($M$) 
for the site centered (in the $S_z=S_{0}$ sector) and Fibonacci trees (in the $S_z=0$ sector). 
However, for the bond centered case (Table~\ref{table:126_bond}), the convergence with $M$ is comparatively slower. 
Based on our data, we conclude that while larger $M$ calculations 
are certainly desirable, the present calculations are reasonable 
and confirm the existence of an anomalous energy scale in the many body spectrum of the bond-centered tree. 
\begin{table}[htpb]
\begin{center}
\begin{tabular}{c||c|c||c|c}
\hline 
$M$ & \multicolumn{2}{c||}{$N_{s}=190$} & \multicolumn{2}{c}{$N_{s}=176$}\tabularnewline
\hline 
 & $E_{GS}$ & $\Delta$ & $E_{GS}$ & $\Delta$\tabularnewline
\hline 
20 & -74.54049387 & 0.9397293 & -76.46983049 & 0.08834668\tabularnewline
60 & -74.54054021 & 0.9397198 & -76.47071767 & 0.08725531\tabularnewline
100 & -74.54054022 & 0.9397198 & -76.47072851 & 0.08725207\tabularnewline
\hline \hline
\end{tabular}
\caption{Energy ($E_{GS}$) and Spin gap ($\Delta$) for the 190 site site-centered and 176 site Fibonacci 
lattices for various values of $M$.}
\label{table:190_176}
\end{center}
\end{table}

The question of finite size scaling can be considered tricky given the small magnitude of the singlet-triplet gap of the bond-centered clusters. 
However, by computing the ground state energy in every $S_{z}$ sector, and fitting the "tower of states" (to be discussed in Sec.~\ref{sec:GS_Tower_of_States}), we believe the errors in estimation of this scaling behavior are mitigated.
\begin{table}[htpb]
\begin{center}
\begin{tabular}{c||c|c|c|c}
\hline 
$M$ & $E_{GS}$ & $\Delta$ & $\left \langle 1 \right | S_{c}^{+} \left| 0 \right \rangle$ & $\left \langle 1 \right | S_{b}^{+} \left| 0 \right \rangle$ \\ \hline 
60 &  -49.3405119 & $3.4\times 10^{-4}$  & 0.310 & 0.370  \\ 
80 &  -49.3412938 & $5.7\times 10^{-4}$ & 0.300 & 0.344 \\ 
100 & -49.3415002 & $6.0\times 10^{-4}$ & 0.280 & 0.337  \\ 
120 & -49.3415347 & $6.0\times 10^{-4}$ & 0.279 & 0.335  \\
140 & -49.3415521 & $6.0\times 10^{-4}$ & 0.278 & 0.335  \\
\hline \hline
\end{tabular}
\caption{Energy ($E_{GS}$), Spin gap ($\Delta$) and $S^{+}$ matrix elements for the central ($c$) and boundary ($b$) sites 
for the 126 site bond-centered for various values of $M$.$\left|0\right\rangle$ and $\left|1\right\rangle$ refer to the lowest singlet and triplet respectively.}
\label{table:126_bond}
\end{center}
\end{table}
 
\section{Ground And Excited States}
\label{sec:GS}
Using the DMRG algorithm presented in section~\ref{sec:DMRG}, we calculate the ground state energy and spin gap and point out the
differences between the bond and site centered clusters. 
To highlight the role of local imbalance, we also compare the excited states of the bond-centered and Fibonacci trees, both of which are globally balanced.
\subsection{Ground State Energy, Spin-spin correlations and Spin Gap}
\label{sec:GS_Energy}
\begin{figure}[htpb]
\includegraphics[width=\linewidth]{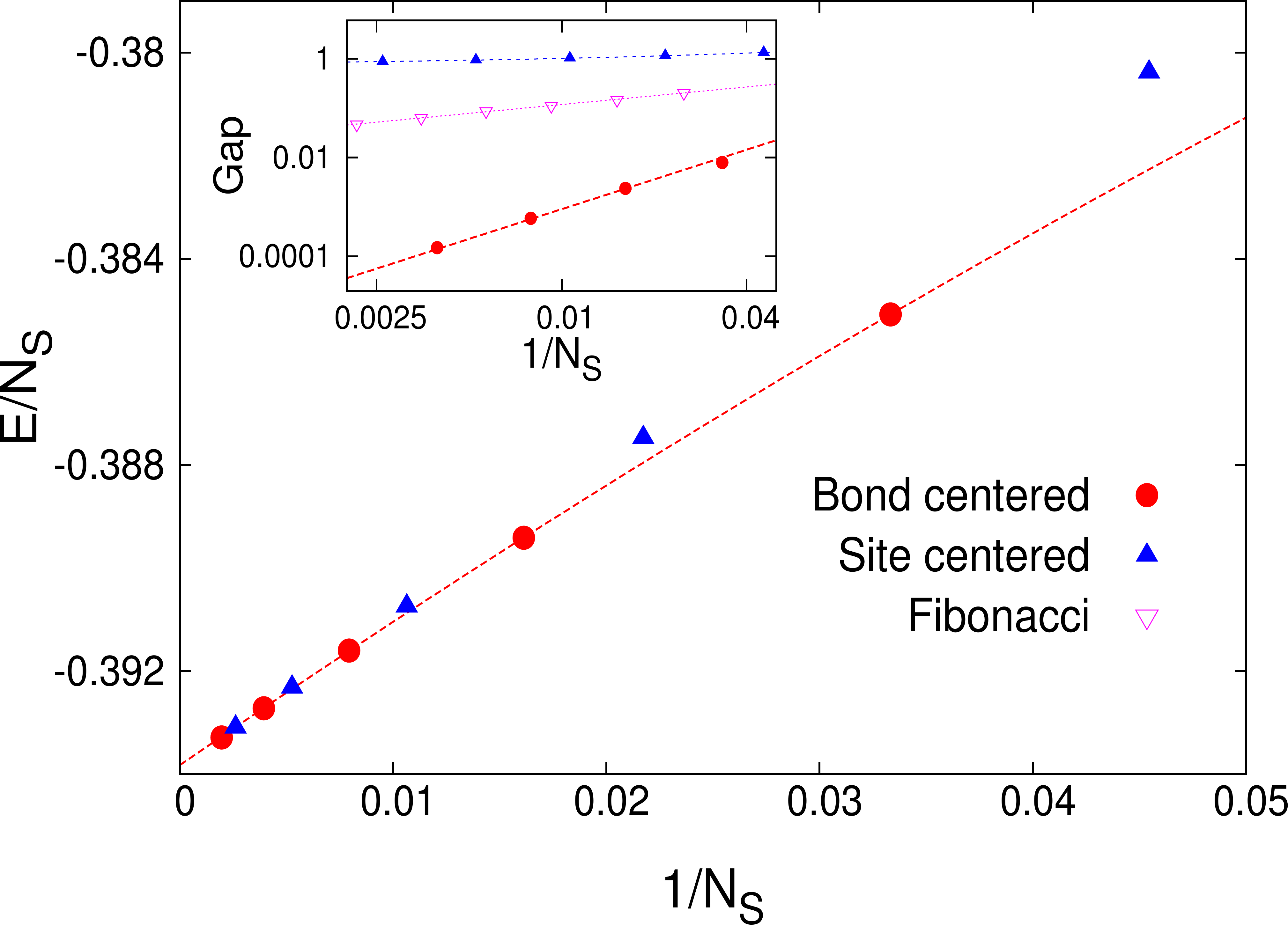}
\caption{Ground state energy per site for the bond centered and site centered Cayley trees 
for various lattice sizes. The Fibonacci Cayley tree energies are out of the scale considered.
A fit to the bond centered results given by equation~\eqref{eq:E_over_N} has been shown.
Inset: Finite size scaling of the energy
gap $\Delta$ plotted on a log-log scale. The bond-centered and Fibonacci clusters appear gapless in the infinite lattice
limit, with a finite size scaling exponent of $\alpha\approx2$ and $\alpha\approx0.6$. However,
the site-centered clusters have a finite $\Delta$ in the infinite lattice 
limit in concordance with the results of Ref.~\onlinecite{Kumar_Ramasesha_Soos}. 
The lines shown are fits to the DMRG data using equations (\ref{eq:clean_gap_site}, \ref{eq:clean_gap_bond}).}
\label{fig:energy_gap}
\end{figure}
We consider the bond-centered, site-centered and Fibonacci clusters and compute the lowest 
energy $E_{GS}(S)$ for various spin sectors $S$. The notation $S_{0}$ will be 
used to refer to the spin of the ground state. 

In order to compute the ground state energy per site in the infinite lattice limit
$e_{0}$ we fit $E_{GS}(S_0)/N_{s}$ to the functional form,
\begin{equation}
\frac{E_{GS}}{N_{s}}=e_{o}+\frac{a}{N_{s}}+\frac{b}{N_{s}^{2}}\label{eq:E_over_N}
\end{equation}
For bond-centered clusters we found $e_{0}=-0.393855(2)$ and for the site-centered clusters $e_{0}=-0.393854(2)$, both of which are consistent
within error bars of extrapolation and with the value
of $e_{0}=-0.39385$ reported for site-centered clusters by Ref.~\onlinecite{Kumar_Ramasesha_Soos}.

In comparison, as Table~\ref{table:energies} shows, the energy per site of 
the Fibonacci tree is significantly lower than either 
of the bond or site centered trees.
This is achieved by the 
formation of very strong nearest neighbor dimers (especially on the boundary, as the degree of dimerization dies down as one proceeds inwards).
The degree of boundary dimerization is more limited in the site and bond centered trees. 

Despite the dissimilarities between the three lattices, 
the "bulk limit" of the estimated energy per bond, based on taking an average of nearest neighbor $\left\langle \vec S_{i} \cdot \vec S_{j} \right\rangle$ over the innermost bonds, 
is roughly identical for all three kinds of Cayley tree, and equals about $ -0.35 J$.   
\begin{center}
\begin{table}[htpb]
\begin{tabular}{c|c|c|c|c|c}
\hline 
Cluster &  $- e_{0}$ & $a$ & $b$ & $\Delta_{+1}$ & $\Delta_{-1}$                         \tabularnewline
\hline 
Bond-cen&  0.393855(2) & 0.29 & $-$1.0 & $\sim N_{s}^{-2}$          & -             \tabularnewline
Site-cen&  0.393854(1) & 0.29 & +0.1 & 0.73 + $2.86/N_{s}^{0.5}$  & $2.19/N_{s}$   \tabularnewline
Fibonacci& 0.435433(1) & 0.17  &  $-$0.4 & $\sim N_{s}^{-0.6}$      & -              \tabularnewline
\end{tabular}
\caption{Ground state energy per site $e_{0}$, finite size scaling parameters for the ground state energy $a,b$ (from equation~\eqref{eq:E_over_N})
and spin gap $\Delta_{+1}$ (from equations~(\ref{eq:clean_gap_site},\ref{eq:clean_gap_bond}) 
for the bond-centered, site-centered and Fibonacci clusters.
We also record the gap $\Delta_{-1} \equiv E_{GS}(S_0-1)-E_{GS}(S_0)$ 
for the site centered cluster.}
\label{table:energies}
\end{table}
\end{center}

We now turn to a discussion of the spin gap. For the site centered tree 
there are two possible spin gaps $\Delta(S_{0}\rightarrow S_{0}\pm 1)$ that can be considered, 
corresponding to $S_{0}\rightarrow S_{0}\pm1$ magnetic transitions (as a shorthand we refer to these gaps as $\Delta_{\pm1}$). 
In the limit of a small magnetic field, for finite system sizes, we get a discrete set of energy levels and the lowest excitation 
involving a single spin flip is $S_{0}\rightarrow S_{0}+1$ transition. 
This excitation $\Delta(S_{0}\rightarrow S_{0}+1)$ has a spin gap in the infinite lattice limit, 
which we obtained by fitting to,
\begin{equation}
\Delta(S_{0}\rightarrow S_{0}+1) = \Delta_{\infty} + \frac{c}{N_{s}^{\alpha}}\label{eq:clean_gap_site}
\end{equation}
and found $\Delta_{\infty}$ to be 0.731(4) and $\alpha \sim 0.5$. 
Interestingly, the spin gap $\Delta(S_{0}\rightarrow S_{0}-1)$ is found to be gapless in the thermodynamic limit 
as seen from the finite-size fit in Table~\ref{table:energies}. 

The bond-centered and Fibonacci clusters appear to be gapless in the infinite lattice limit, 
based on cluster sizes up to 254 and 464 sites respectively. 
We computed the finite size scaling of this gap using,
\begin{equation}
\Delta(0\rightarrow 1)\sim N_{s}^{-\alpha}\label{eq:clean_gap_bond}
\end{equation}
Empirically, the value of $\alpha\sim 0.6$ for Fibonacci and $\alpha\sim2$ for the bond centered clusters matches 
rather well with our data (see inset of Fig.~\ref{fig:energy_gap}).

Denoting the ground state as $\left| 0 \right\rangle$, we also compute the ground state connected correlation function defined here as,
\begin{equation}
G_{ij}\equiv\left\langle 0\left|\vec S_{i} \cdot \vec S_{j}\right|0\right\rangle -\left\langle 0\left|S_{i}^{z}\right|0\right\rangle \left\langle 0\left|S_{j}^{z}\right|0\right\rangle \label{eq:G_ij_general}
\end{equation}
For the bond-centered and Fibonacci clusters, $\left\langle 0\left|S_{i}^{z}\right|0\right\rangle =0$
for all $i$, and hence we have,
\begin{equation}
G_{ij}=\left\langle 0\left|\vec S_{i} \cdot \vec S_{j}\right|0\right\rangle \label{eq:G_ij_balanced}
\end{equation}

\begin{figure}[htpb]
\includegraphics[width=\linewidth]{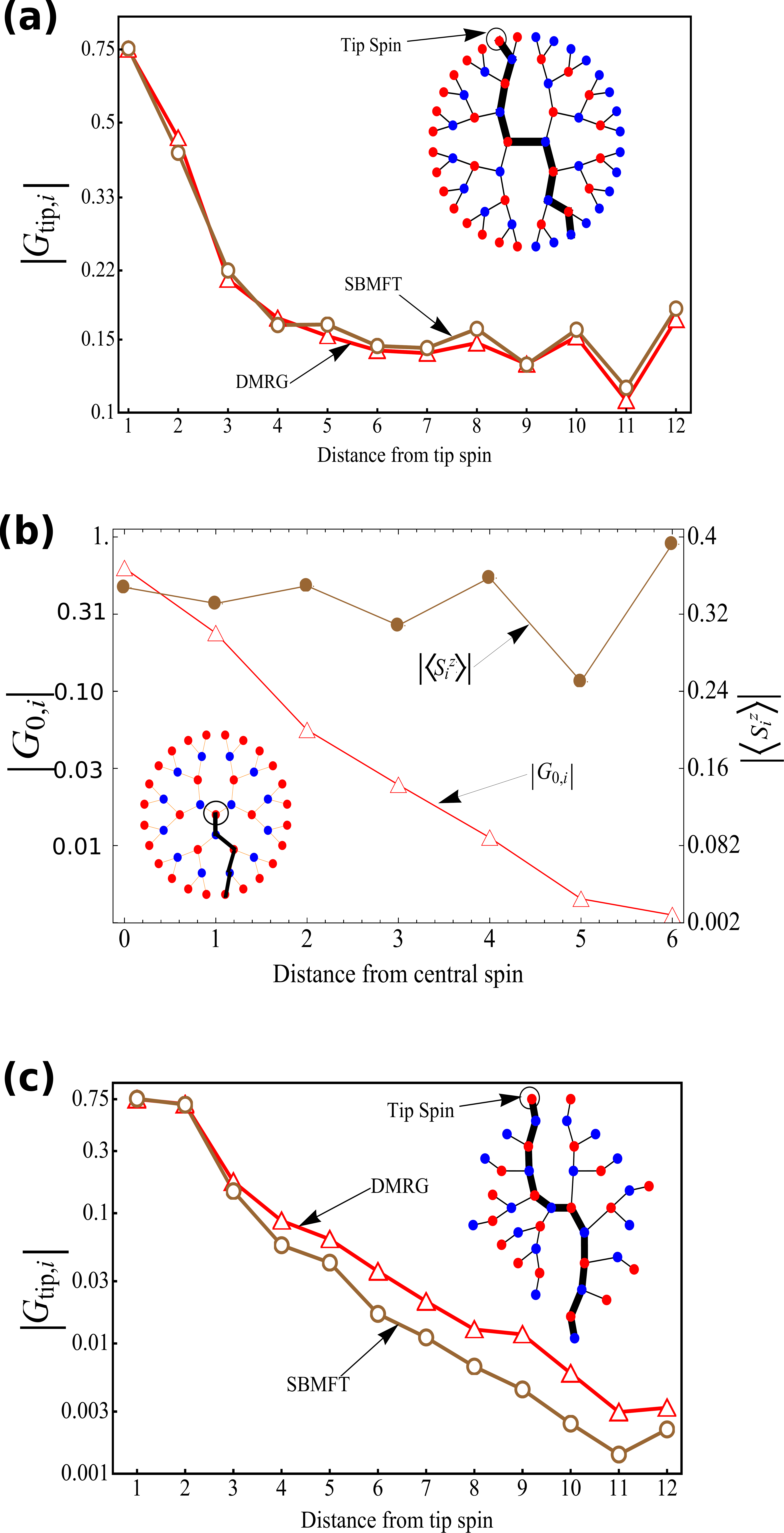}
\caption{(Color online) 
Ground state spin-spin correlations $G_{a,i}$, as in equation (\ref{eq:G_ij_general}),
for the three kinds of Cayley tree. 
One (reference) spin $a$ is held fixed while the other spin $i$ is take at 
distance $a$ along the highlighted path. 
In (a) and (c), the reference spin is at a tip ($a \to \text{tip}$), and
DMRG results are compared with numerical solutions
of Schwinger Boson mean field theory (SBMFT) 
from Section~\ref{sec:SBMFT}.
In (b), the reference spin is at the central (``root'') site ($a \to 0$).
(a) Bond-centered tree  with $N_{s}=126$ sites.
The SBMFT corrrelations shown have been scaled up by an
overall factor of $1.8$ to take into account corrections beyond mean field (in the broken symmetry phase).
The DMRG and SBMFT results show good agreement, asymptoting to a constant.
(b) Site-centered cluster with $N_{s}=190$ sites,
in the maximum $S_z$ member of the ground state multiplet
($S_z=S_0$).
The magnetization $|\left\langle S_{i}^{z} \right\rangle|$ is also shown,
as a function of distance from the center. 
Even though the connected correlation function decays to zero exponentially fast, 
the long range order is encoded in the fact that that the magnetization is non zero.
(c) Fibonacci tree with $N_{s}=40$ sites.
For the "quantum disordered" phase, the SBMFT correlations were scaled up by an overall factor 
of 3/2 
(For details see section~\ref{sec:SBMFT_Corr}).
Correlations appear to be decaying exponentially with distance.}
\label{fig:sample_corrs}
\end{figure}

Fig.~\ref{fig:sample_corrs} shows some sample correlation functions on all three lattices.
The marked difference in the behavior of the spin gap and the spin correlations
between the site- and bond-centered clusters can be attributed to a different
manifestation on the two type of clusters of the spontaneous spin symmetry breaking 
occurring in the thermodynamic limit. First, the behavior of the spin correlations can
be understood in the following way: on the site centered clusters, the system has
an extensive total spin $S=2^{g-1}$ in the ground state. By choosing a particular state
out of this large multiplet it is possible to orient the N\'eel vector at no energy cost in 
this finite size system. In particular if one considers the $S^z=S$ state of the multiplet, the
local $\langle S^z_i\rangle$ expectation values will reflect the ordering pattern directly.
This situation is somewhat analogous to ferrimagnetic systems. In the case of the bond-centered
clusters we have a unique $S=0$ ground state, which forbids finite  $\langle S^z_i\rangle$ expectation
values on a finite system, and the long-range order then has to be coded in the correlation functions
leveling off to a finite value at long distances. 

\subsection{Low energy Tower of States}
\label{sec:GS_Tower_of_States}
For the balanced Heisenberg antiferromagnet ($n_{A}=n_{B}$ with a singlet ground state) 
on a regular unfrustrated lattice (e.g. square in 2D or cubic in 3D), with number of sites $N_{s}$, 
it has been noted and argued~\cite{Ziman,Gross}
that the low energy Hamiltonian 
can be described by a rigid rotor model, 
\begin{equation}
H_{rot} = \frac{{\vec S}^2}{2I} 
= \frac {S(S+1)}{2I} \label{eq:gap_scaling}
\end{equation}
where ${\vec S}$ it the total angular momentum (spin), and
\begin{equation}
I \cong \chi N_{s}\label{eq:M_chi}
\end{equation}
where $\chi$ is the susceptibility of $\vec S$ to a field coupling to it.
Thus, we have a sequence of multiplets, 
popularly referred to as the {}"Anderson Tower of States",
which become degenerate in the limit $N_s \to \infty$ thus
making $SU(2)$ symmetry breaking possible in this limit. However, contrary to the effects of spontaneous spin rotational symmetry breaking on regular lattices, the Cayley tree does not have any Goldstone modes~\cite{Sondhi}.

To observe the "Anderson tower of states" on the Cayley tree,
we compute the ground state energy in every $S_{z}$ sector. 
This may be identified with the multiplet energy $E(S)$,
since $E(S)$ is monotonic in $S$ and $S\ge S_z$.

\begin{figure}[htpb]
\includegraphics[width=\linewidth]{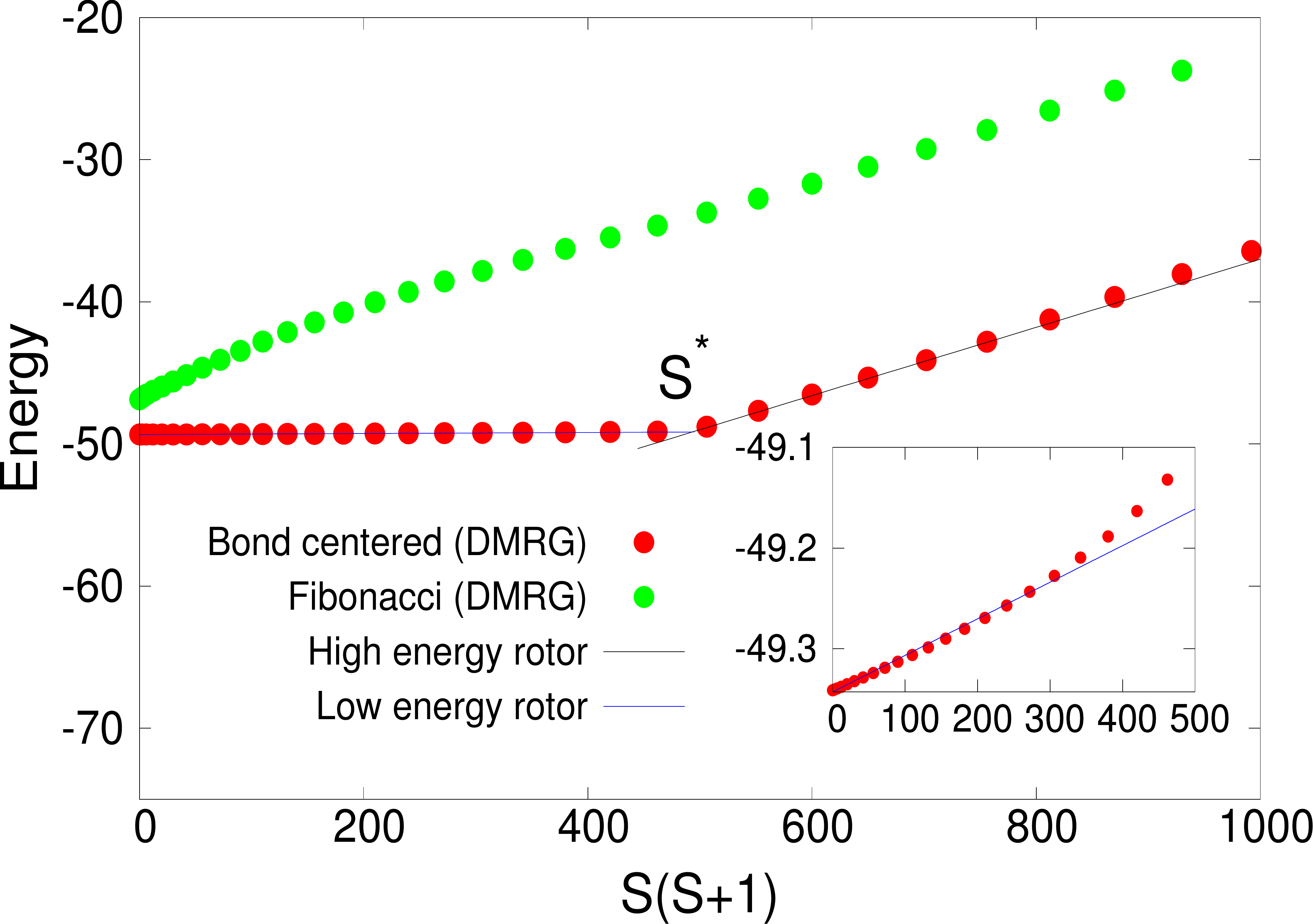}
\caption{Lowest energy level in every $S_{z}$ sector for the 108 Fibonacci and the 126 site bond-centered Cayley
tree. The range of $S$ from 0 to $S^{*}$ has been magnified and shown in the inset for the 126 site cluster. 
It shows a tower of states with a much larger moment of
inertia than expected from the Anderson tower. This is seen as a sharp kink like feature at $S^{*}$. In contrast, for the Fibonacci tree, 
the transition from the low to high $S$ behavior is less well defined.
}
\label{fig:rotorspectrum}
\end{figure}

For the bond centered clusters, a tower of states exists, but
an anomalous one.
In Fig.~\ref{fig:rotorspectrum} we observe that in the bond-centered case,
the $E(S)$ curve consists of {\it two} linear pieces, joined at a 
critical spin $S^*$ which depends on the cluster size.
In effect, the system has two moments of inertia,
$I_{\text{low}}$ for $S<S^{*}$ 
and the (much smaller) $I_{\text{high}}$ for $S>S^{*}$.
Finite size fits of the moment of inertia show that $I_{\text{low}} \approx 0.191 N_{s}^{2}$ while $I_{\text{high}} \approx 0.358 N_{s}$;
it will be our task in Sec.~\ref{sec:SMA_Giant_Spin}, to explain this difference.
We also observe that 
\begin{equation}
S^{*}=2I_{b}
\end{equation}
where $I_{b}$ is the sublattice imbalance 
on one half of the bond-centered cluster
as defined in equation~\eqref{eq:Imbalance_bond_centered}.

For the Fibonacci tree, we do not find a clear distinction between the two linear pieces as seen in Fig.~\ref{fig:rotorspectrum}. 
The scaling of the energy gaps ($E(S+1) - E(S)$) changes from $N_{s}^{-0.6}$ for small $S$, to the $1/N_{s}$ 
Anderson scaling for large $S$.

In contrast with the gapless spectrum of the bond-centered and Fibonacci clusters, 
the site-centered case has a finite spin gap (see Table~\ref{table:energies}) in the infinite lattice limit.

\begin{figure}[htpb]
\includegraphics[width=0.9\linewidth]{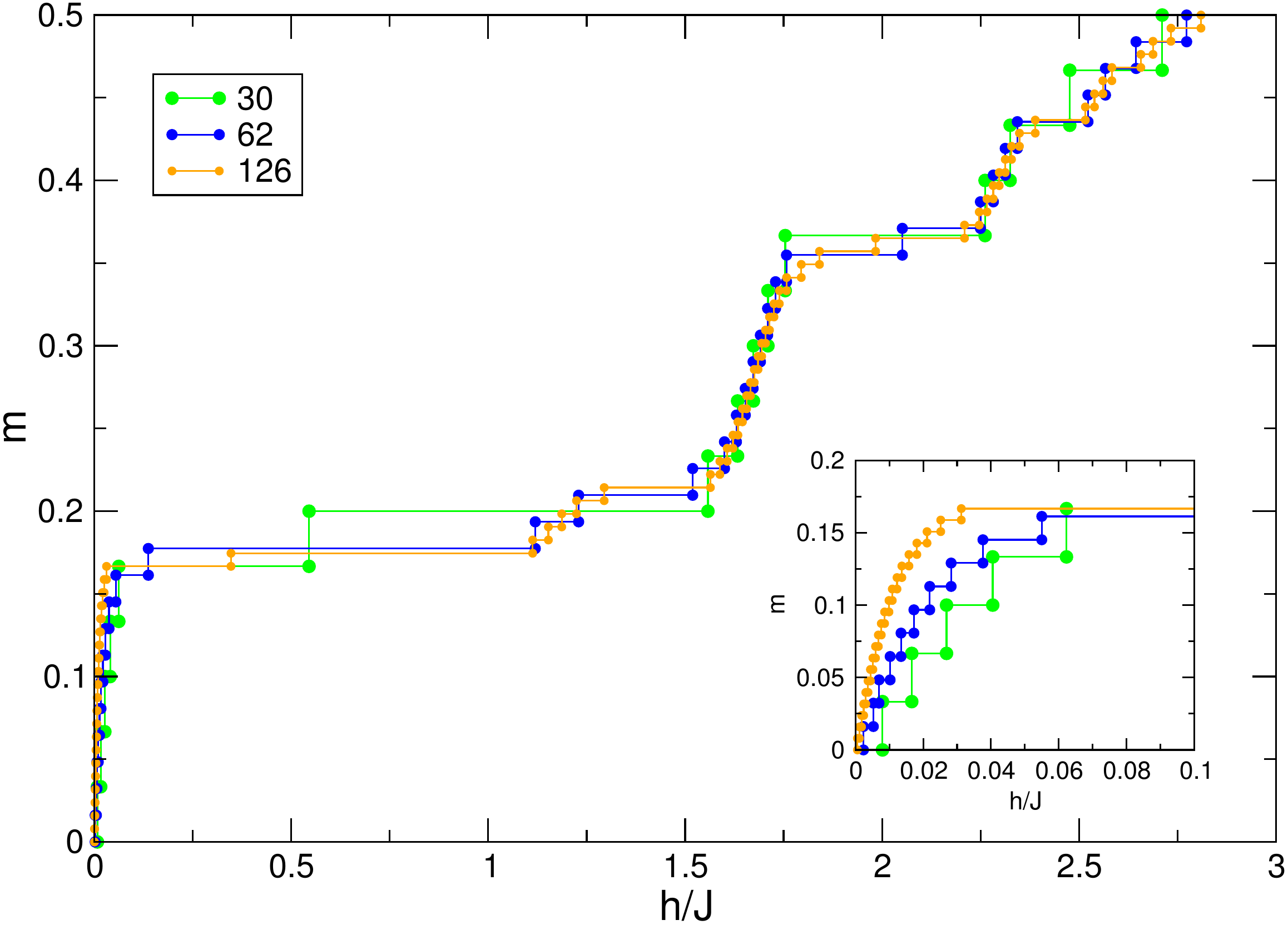}
\caption{Magnetization curves for bond-centered Cayley trees of various sizes obtained
using DMRG. Inset:
The rapid rise of the magnetization with application of a small magnetic
field indicates a susceptibility diverging with system size. }
\label{fig:m_h}
\end{figure}

A complementary probe of the low energy physics is the magnetization ($m$) defined as,
\begin{equation}
m=\frac{1}{N_{s}} \sum_{i}\langle S_{i}^{z}\rangle_{GS}
\label{eq:mag definition}
\end{equation}
as a function of a uniform applied magnetic field $h$. For the bond-centered clusters (see Fig.~\ref{fig:m_h}), we observe a rapid increase in
magnetization for small $h$ and it seems the saturation magnetization 
is about $m^* \sim 1/6$ (i.e. $m^*/m_\mathrm{sat} \sim1/3$). 
Beyond this rapid rise of the magnetization at small field, the magnetization curve displays a surprisingly
rich structure, with several plateau-like features at intermediate magnetization, linked by more susceptible
parts of the magnetization curve. We note that the first plateau at $m^*$ seems to have a similar extent
in magnetic field as for the site-centered clusters studied in Ref.~\onlinecite{Kumar_Ramasesha_Soos}. 
The detailed characterization of the magnetization curve as $N_{s}\rightarrow\infty$ appears to be an 
interesting problem for future studies. 

\begin{figure}[htpb]
\includegraphics[width=0.9\linewidth]{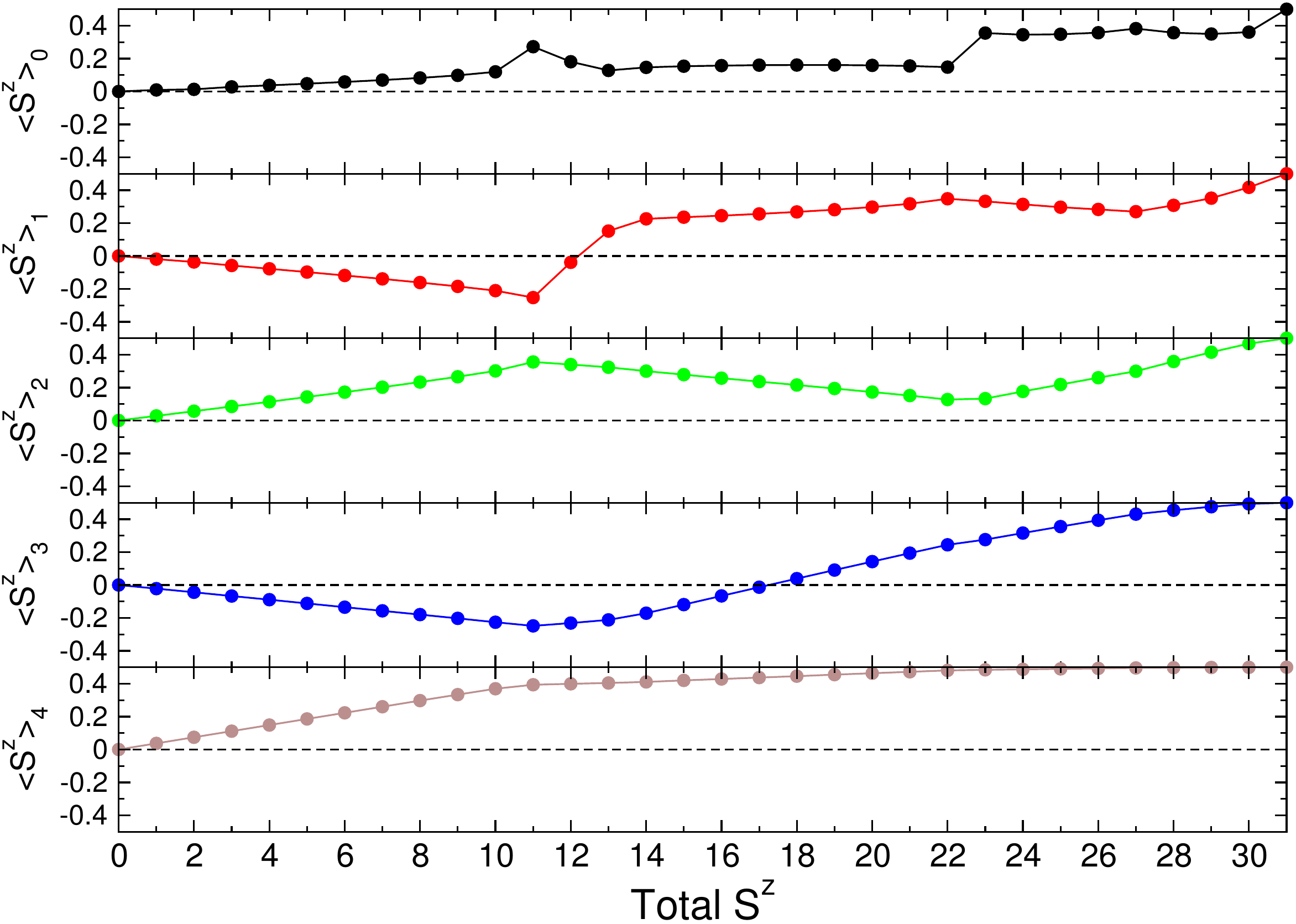}
\caption{Magnetization curves for sites on various shells of the 62 site bond-centered
Cayley tree. The subscript refers to the shell on which the site is
present, that is, 0 refers to the central two sites and the 4 refers to the
boundary.} 
\label{fig:local_m}
\end{figure}

In Fig.~\ref{fig:local_m}, we also show the magnetization curves for sites
on various shells of the 62-site bond-centered Cayley tree. For
small Sz (or equivalently small magnetic fields), we infer that
while the boundary spins are most susceptible, the spins in
the interior also have a comparably high susceptibility, which
decreases as we go inwards.

\section{Single Mode Approximation for the excited state}
\label{sec:SMA}
Can we understand the origin of the {}"anomalous" states in the low energy spectrum for the bond-centered tree?
In order to address this question, we study the nature of the triplet excited state 
and its relation to the ground state using the single mode approximation~\cite{Feynman}(SMA for short). 
Assuming the knowledge of the ground state wavefunction
$\left|0\right\rangle$ (analytically or from a numerical method), the SMA ansatz for the trial state 
$\left|1'\right\rangle \equiv \left|S=1,S_{z}=1\right \rangle$ state is given by,
\begin{equation}
\left|1'\right\rangle =\frac{1}{\mathcal{N}_{1'}}\sum_{j=1}^{N_{s}} u_{j}S_{j}^{+}\left|0\right\rangle \label{eq:SMA_ansatz}
\end{equation}
where $u_{j}$ are variational parameters and $\mathcal{N}_{1'}$ is a normalization factor given by,
\begin{equation}
\mathcal{N}_{1'}\equiv\sqrt{\sum_{k,l}u_{k}^{*}u_{l}\left\langle 0\left|S_{k}^{-}S_{l}^{+}\right|0\right\rangle}\label{eq:norm_1}
\end{equation} 

Using the spin symmetry of the Hamiltonian (and hence
its eigenfunctions) and the fact that the ground state has $S_{z}=0$,
the normalization factor $\mathcal{N}_{1'}$ (\ref{eq:norm_1}) can be written as,
\begin{equation}
\mathcal{N}_{1'}=\sqrt{\sum_{k,l}\frac{2}{3}u_{k}u_{l}G_{kl}} \label{eq:norm_2}
\end{equation}
where $G_{kl}$ is the spin-spin correlation function previously defined in equation~\eqref{eq:G_ij_balanced}. 

For a singlet ground state, observe that, there is a gauge degree of freedom in the choice of the SMA wavefunction (\ref{eq:SMA_ansatz}),
an arbitrary constant shift $u$,
i.e.,
\begin{equation}
\sum_{i}u_{i} S_{i}^{+}\left|0\right\rangle = \sum_{i}(u_{i}+u) S_{i}^{+}\left|0\right\rangle \label{eq:gauge_symmetry}
\end{equation}
since for a ground state with total spin $S=0$,
\begin{equation}
u\sum_{i}S_{i}^{+}\left|0\right\rangle =uS_{tot}^{+}\left|0\right\rangle =0\label{eq:S_tot_plus}
\end{equation}
It can also be shown~\cite{FN_SMA} that the normalization $\mathcal{N}_{1'}$ in equation~\eqref{eq:norm_2} is invariant under the transformation $u_{i}\rightarrow u_{i}+u$.

For a given trial state $\psi_{T}$ which is a function
of some parameters $u_{j}$, the variational principle guarantees that,
\begin{equation}
E_{T}\equiv\frac{\left\langle \psi_{T}\left(\left\{ u_{j}\right\} \right)\left|H\right|\psi_{T}\left(\left\{ u_{j}\right\} \right)\right\rangle }{\left\langle \psi_{T}\left(\left\{ u_{j}\right\} \right)\left|\psi_{T}\left(\left\{ u_{j}\right\} \right)\right.\right\rangle }
\end{equation}
where $E_{0}$ refers to the energy of the lowest lying state with the same symmetry as the trial wavefunction. The best wavefunction is obtained by optimization of the parameters
$\{u_j\}$ by minimizing the variational energy $E_{T}$.
Note that within the SMA formalism, the ground state (and hence the ground state
energy) is assumed, which implies that the SMA spin gap is also a variational
estimate for the true spin gap. 

Here we will show that the SMA does turn out to be a very good ansatz for this system based
on the close to 100~\% overlap of the SMA wavefunction with the wavefunction from DMRG. Our procedure does not require explicit knowledge of the wavefunction, rather only certain matrix elements and correlation functions are necessary. We will derive our intuition from numerical calculations, we construct coefficients $u_{j}$ occurring in equation~\eqref{eq:SMA_ansatz} to obtain a simple variational state with a gap that goes to 0 faster
than $1/N_{s}$. The aim of this section is thus to understand the operator that creates the triplet wavefunction from the ground state singlet. 
This in turn will tell us how the spins collectively act, which will be used to understand the existence of an {}"anomalous" energy scale.

A similar SMA calculation was performed by Laumann et al.~\cite{Laumann_AKLT}, who considered AKLT models on the Cayley tree, where the analytical form of the correlation functions is known.
Instead, we use the values of $\left\langle 1 \right|S_{i}^{+}\left|0\right\rangle$
and $G_{ij}$ from our DMRG calculations as inputs for our analysis. In addition, we make no assumptions about the variational parameters $\{u_{j}\}$.
  
\subsection{Obtaining the SMA coefficients from maximization of overlap
with the true wavefunction}
\label{sec:SMA_optimize_overlap}

The overlap of an approximate wavefunction with the exact one can serve as a good measure of its quality.
Thus we consider the overlap of the SMA wavefunction with the true triplet state
$\left| 1 \right\rangle \equiv \left| S=1,S_{z}=1 \right\rangle$ i.e.,
\begin{eqnarray}
O & \equiv & \left\langle 1\left|1'\right.\right\rangle \nonumber \\
 &=&\frac{\sum_{j}u_{j}f_{j}}{\sqrt{\sum_{k,l}\frac{2}{3}u_{k}u_{l}G_{kl}}}\label{eq:overlap_sz_sz}
\end{eqnarray}
and try to maximize it. We have defined $f_{i}$ to be,
\begin{equation}
f_{i}\equiv\left\langle 1\left|S_{j}^{+}\right|0\right\rangle \label{eq:c_i_defn}
\end{equation}
We christen this term the {}"flippability" of a spin. 
This is motivated from its very definition: 
the more easily flipped spins have a larger contribution to the formation of the first excited triplet.

Now we present a method to obtain the optimal parameters $u_{j}$ to construct $\left|1'\right\rangle$. 
To meet the requirements of a high overlap of the SMA wavefunction $\left|1'\right\rangle$ 
with the exact many body triplet $\left|1\right\rangle$
, subject to the constraint that it is normalized, we devise a cost function $C_{SMA}$ defined as,
\begin{equation}
C_{SMA} = -\sum_{j}u_{j}f_{j}+\lambda\left(\mathcal{N}^{2}_{1'}-1\right)\label{eq:Cost_overlap}
\end{equation}
where we have introduced $\lambda$ as a Lagrange multiplier. Taking
$\{u_{j}\}$ as our variational parameters, and setting the derivatives
of $C_{SMA}$ to 0, we get,
\begin{eqnarray}
\frac{\partial C_{SMA}}{\partial u_{i}} = -f_{i}+\frac{4}{3}\lambda\sum_{l\neq i}u_{l}G_{il}+\lambda u_{i} = 0 \label{eq:Derivative_cost}
\end{eqnarray}
Thus we get the set of equations (one equation for each $i$) ,
\begin{equation}
f_{i}=\frac{4}{3}\lambda\sum_{j}G_{ij}u_{j}\label{eq:c_4_G_u}
\end{equation}
To explicitly obtain a state $\left|1'\right\rangle $ which has a high
overlap with $\left|1\right\rangle $ , we solve the above linear
equations for $u_{i}$ numerically. Note that the matrix $G$ always has exactly
one zero eigenvalue because of the gauge degree of freedom (\ref{eq:gauge_symmetry}). Hence we can not simply invert $G$ to obtain
$u_{i}$; instead, we directly solve the linear system of equations (\ref{eq:c_4_G_u}) using dgesv in LAPACK.

A natural choice of gauge for the parameters $\{u_{j}\}$ is to satisfy,
\begin{equation}
\sum_{i}u_{i}=0\label{eq:gauge_condition}
\end{equation}

\begin{figure}[htpb]
\centering
\includegraphics[width=\linewidth]{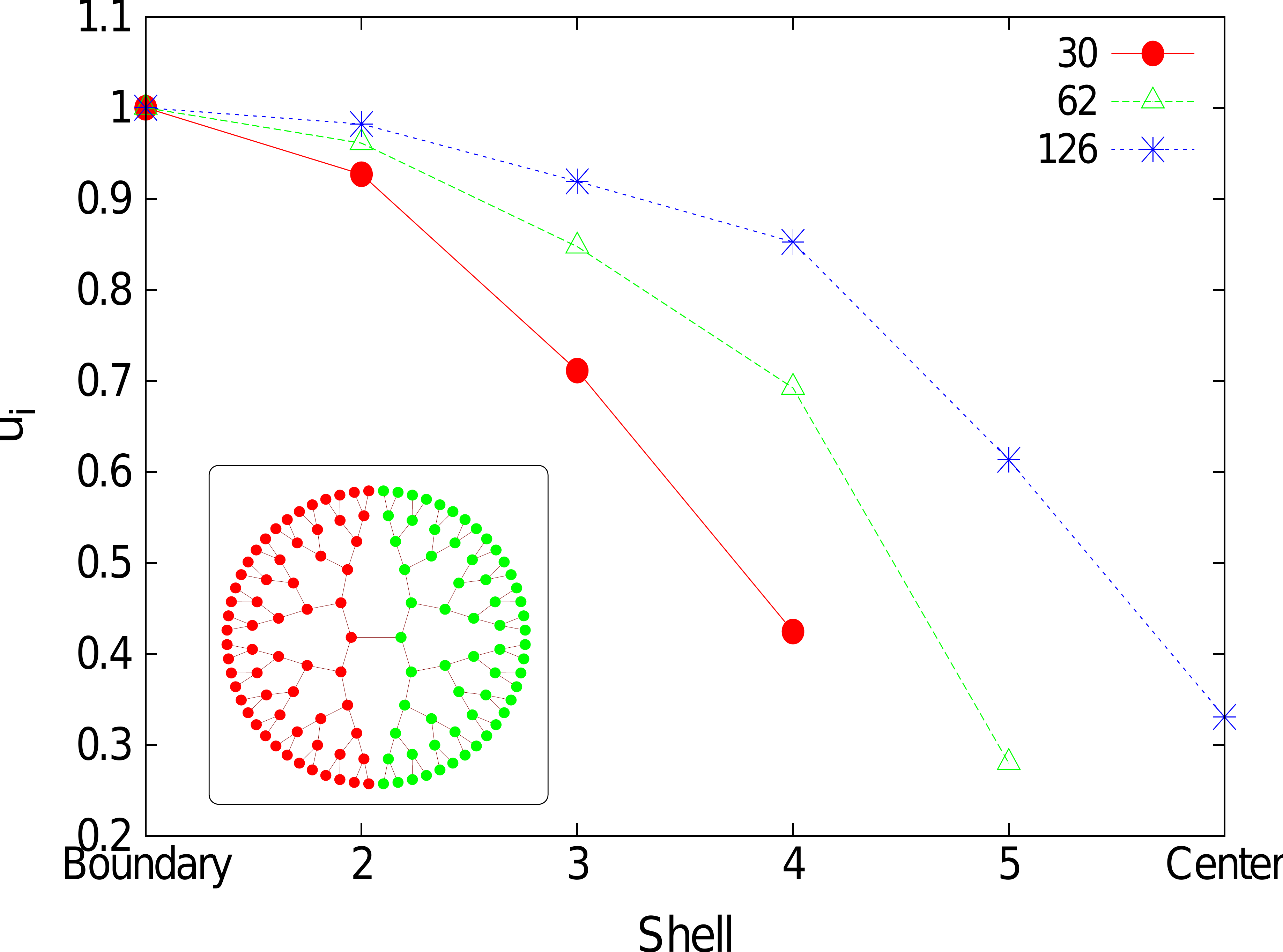}
\caption{(Color online) Amplitude of the SMA coefficients $u_{i}$ for various shells (normalized with respect to amplitude on the boundary) 
for the $N_{s}=$ 30,62 and 126 site bond-centered lattices.
Inset: The sign structure of $u_{i}$ is the same as equation (\ref{eq:SMA_choice}).
Dark(red) and light(green) indicates negative and positive $u_{i}$ respectively.}
\label{fig:optimal_ui}
\end{figure}

Our observation from the numerical solution of equation (\ref{eq:c_4_G_u}) for the bond-centered Cayley
tree is that $u_{i}$ $>0$ for $i$ on one side of the central bond and $u_{i}<0$
on the other side. We have also plotted the amplitudes of the optimal
SMA coefficients for the 30, 62 and 126 site bond-centered Cayley trees
in Fig.~\ref{fig:optimal_ui}.

\subsection{Comparison of various SMA wavefunctions}
\label{sec:SMA_wf_comparison}

We next try to understand the qualitative nature of the SMA solution
from the perspective of minimizing the triplet energy. We
consider various functional forms for $u_{i}$ and numerically compute 
their overlap with the exact triplet and compare SMA gap estimates.

For the nearest neighbor Heisenberg model,
the SMA gap is given by (for a derivation refer to Appendix~\ref{sec:derivSMA}),
\begin{equation}
\Delta_{SMA}=\frac{-\sum_{\left\langle k,l \right\rangle}\left(u_{k}-u_{l}\right)^{2}G_{kl}}{2\sum_{ij}u_{i}u_{j}G_{ij}}\label{eq:delta_sma_Gkl}
\end{equation}
Observe that the numerator and denominator (being proportional to $\mathcal{N}_{1'}^{2}$) 
are invariant under the transformation $u_{i}\rightarrow u_{i}+u$.

To minimize the SMA gap, one would like to minimize the numerator
and maximize the denominator of equation (\ref{eq:delta_sma_Gkl})
(note both the numerator and denominator are positive). To minimize
the numerator, we can try to keep $u_{k}\approx u_{l}$ for as many
bonds as possible, and hence consider the {}"left-right" ansatz,
\begin{equation}
u_{i} = \begin{cases}+1\;\;\;\;\;\;\;\; i\in\mbox{left of central bond} \\
 -1\;\;\;\;\;\;\;\; i\in\mbox{right of central bond}\label{eq:SMA_choice}
\end{cases}
\end{equation}
which is consistent with the gauge condition (\ref{eq:gauge_condition}). This
sign structure is in concordance with the numerical solution of equations
(\ref{eq:c_4_G_u}) for the bond centered Cayley tree.

Note that this is quite contrary to the {}"staggered pattern",
one obtains by solving equations (\ref{eq:c_4_G_u}) for the square
lattice. The staggered pattern is defined as,
\begin{equation}
u_{i} = \begin{cases}+1\;\;\;\;\;\;\;\; i\in\mbox{even sublattice} \\
 -1\;\;\;\;\;\;\;\; i\in\mbox{odd sublattice}\label{eq:SMA_staggered_choice}
\end{cases}
\end{equation}
The staggered pattern is an energetically expensive solution for the
bond-centered Cayley tree. Even though it maximizes the denominator
making it $O(N_{s}^{2})$, the numerator is also large i.e. $O(N_{s})$.
Thus the SMA gap scales only as $O(1/N_{s})$

Table~\ref{table:overlap_SMA} verifies the arguments presented above by explicitly
listing out the SMA gap and overlap with the exact wavefunction for
the various choices of $u_{i}$ we have considered here. Our inference is that the optimal and the {}``left-right''
ansatz are qualitatively similar and yield a much smaller SMA gap than the "staggered" ansatz.

The SMA calculations suggest that all the spins are involved in the construction of the first excited state from the ground state. 
The $u_i$ corresponding to the central spins is roughly a third of the $u_i$ of the boundary spins in the "optimal solution" 
and exactly as much as the boundary spins in the "left-right" ansatz. The point to note here is that in either case the contribution of the spins in the interior is not small. 
This suggests that the antiferromagnetic bonds between successive shells do a reasonable job of locking spins 
together (within each half of the bond centered tree), 
resulting in an emergent degree of freedom which is what we call a "giant spin". 
This interpretation will be established next in section~\ref{sec:SMA_Giant_Spin}. 
\begin{table}[htpb]
\begin{center}
\begin{tabular}{|>{\centering}p{1cm}|>{\centering}p{2.4cm}|>{\centering}p{2.4cm}|>{\centering}p{2.4cm}|}
\hline 
$N_{s}$ & $u_{i}$ &  SMA Gap & Overlap\tabularnewline
\hline 
 & Optimal & 0.0135 & 0.998\tabularnewline
\cline{2-4} 
30 & Left-Right & 0.0341 & 0.993\tabularnewline
\cline{2-4} 
 & Staggered & 0.2680 & 0.912\tabularnewline
\hline 
 & Optimal & 0.0039 & 0.999\tabularnewline
\cline{2-4} 
62 & Left-Right & 0.0116 & 0.997\tabularnewline
\cline{2-4} 
 & Staggered & 0.1612 & 0.946\tabularnewline
\hline 
 & Optimal & 0.0010 & $\approx1$\tabularnewline
\cline{2-4} 
126 & Left-Right & 0.0028 & $\approx1$\tabularnewline
\cline{2-4} 
 & Staggered & 0.0905 & 0.975\tabularnewline
\hline 
\end{tabular}
\caption{SMA gap and wavefunction overlap with excited state from DMRG for various functional forms of $u_{i}$}
\label{table:overlap_SMA}
\end{center}
\end{table}

\subsection{The {}"Giant Spins" Picture}
\label{sec:SMA_Giant_Spin}
\begin{figure}[htpb]
\includegraphics[width=0.95\linewidth]{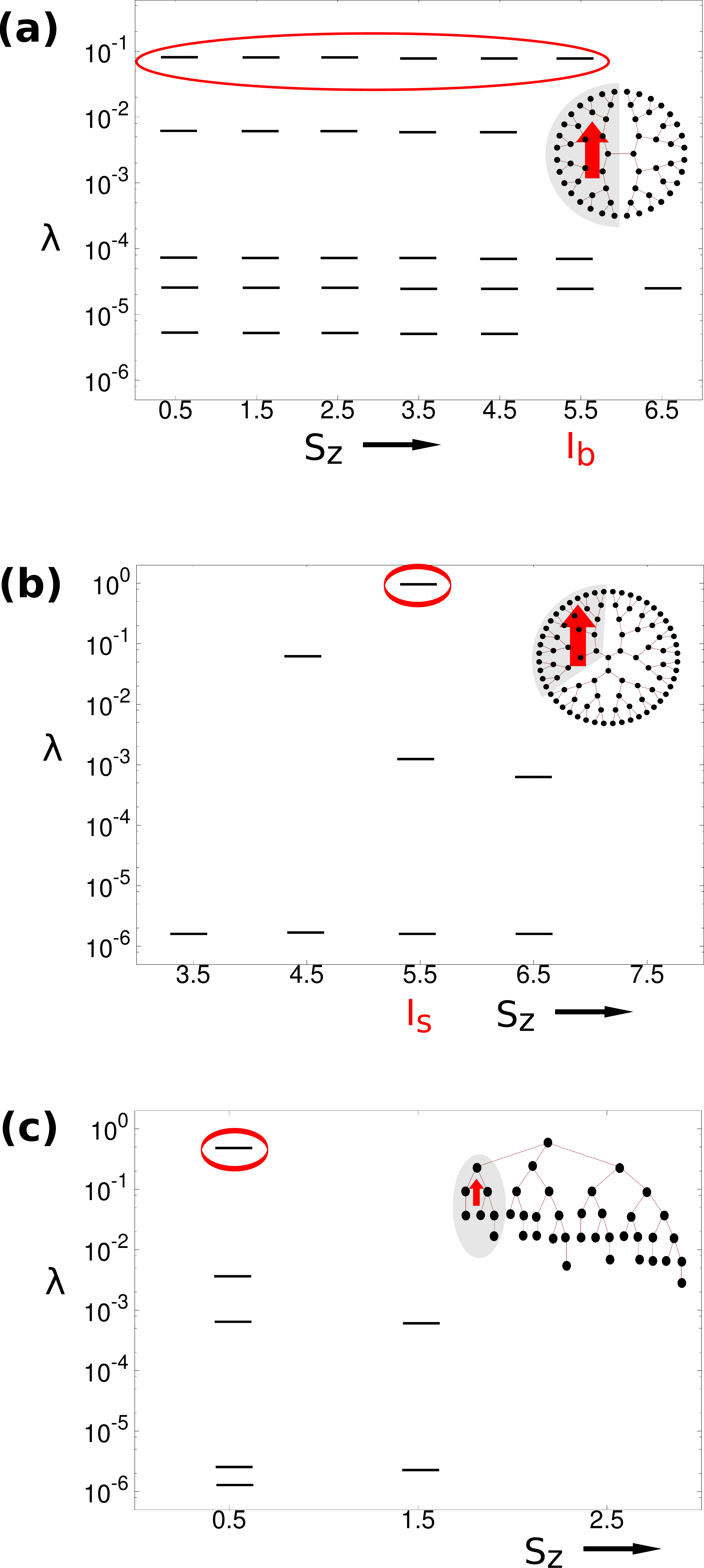}
\caption{(Color online) The entanglement spectrum for the bond-centered,site-centered and Fibonacci trees as shown in the inset. 
$\lambda$ refers to the eigenvalue of the reduced density matrix of the shaded area. 
The ground state for the bond-centered and Fibonacci clusters was a singlet and only the $S_z>0$ sectors are shown. 
For the site-centered cluster we chose to work with the maximal $S_z$ sector (which is the $S_z$=16 for the 94 site cluster). 
$I_{b}$ and $I_{s}$ denote the {}"imbalance" metric defined in the introduction (refer to equation~\eqref{eq:Imbalance_bond_centered}).
For the bond-centered case, the largest degenerate eigenvalues of the reduced density matrix indicate a multiplet whose spin length exactly equals the imbalance $I_{b}$. For the site-centered case, the density matrix has largest weight in a state whose spin is $I_{s}$.
For the Fibonacci case, a spin 1/2 state has the largest weight in the density matrix.}
\label{fig:dm_cuts}
\end{figure}
As we inferred previously, there are indications that strong antiferromagnetic correlations
force all spins in one half of the bond-centered cluster to act collectively as a single magnetic moment.
We make this understanding more concrete in the present section. 

We divide the bond-centered Cayley tree into two equal halves at the central bond.
Using the ground state, we compute the reduced density matrix $\rho_{RDM}$ (see equation~\eqref{eq:RDM}) 
of one of the halves and diagonalize it. 
The corresponding eigenvalues are arranged by total $S_{z}$ and the resultant plot is the {}"entanglement spectrum". 
Appropriate cuts are also chosen for the site-centered and Fibonacci trees as shown in Fig.~\ref{fig:dm_cuts}.   

The entanglement spectrum shows a copy of the largest eigenvalue in every $S_{z}$ sector ranging from $-I_{b}$ to $+I_{b}$, 
where $I_{b}$ is the net sublattice imbalance and is given by $(2^{g}\pm1)/6$ as mentioned in equation~\eqref{eq:Imbalance_bond_centered}. 
This indicates the presence of a {}"giant spin" of spin length $I_{b}$ whose multiplet is given by the 
eigenvectors corresponding to these large eigenvalues. 
Given this picture, we explain the existence of the {}"anomalous" scale of energies.

\begin{figure}[htpb]
\includegraphics[width=\linewidth]{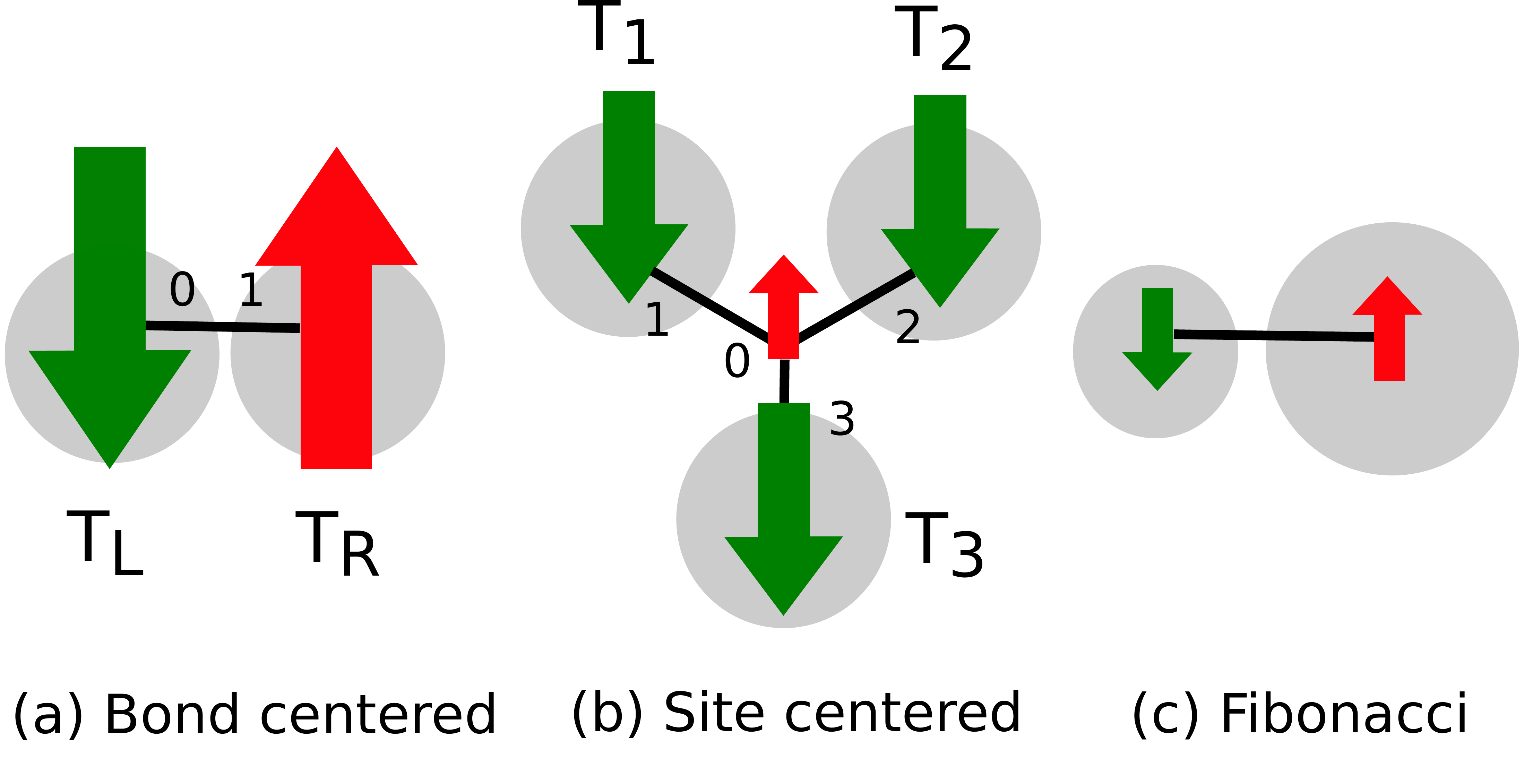}
\caption{(Color online) Schematic of the "giant spins", which are
the low-energy degrees of freedom for the bond- and site centered
clusters (and which are absent for the Fibonacci tree). The numbering
of sites shown here is used for the purpose of explaining our arguments
in the text.}
\label{fig:giant_spins}
\end{figure}

\subsubsection{Bond-centered tree}
The Heisenberg Hamiltonian on the bond centered Cayley tree may be rewritten as,  
\begin{equation}
H=H_{\text{left}}+H_{\text{right}}+J\vec S_{0}\cdot\vec S_{1} \label{eq:Heisenberg_Hamiltonian_bond_centered}
\end{equation}
where 0 and 1 refer to the central two sites of the tree (as has
been schematically represented in Fig.~\ref{fig:giant_spins}).
We treat the term corresponding to the central bond $J\vec S_{0}\cdot\vec S_{1}$ as a perturbation within degenerate first order perturbation theory. 
The many body ground state on each half is a degenerate multiplet of spin $I_{b}$. 
Since all spins on the left and right contribute in a definite proportion to the {}"giant spin" operators $\mathbf{{T}}_{L}$ and $\mathbf{{T}}_{R}$, 
one can re-express the expectation values of the $J\vec S_{0}\cdot\vec S_{1} $ in terms of $\mathbf{{T}}_{L}$ and  $\mathbf{{T}}_{R}$. 
Note that expectation values of the term  $J\vec S_{0} \cdot \vec S_{1}$ are computed in the product basis of the two systems given by $\left|T_{L},T_{L}^{z}\right\rangle \otimes\left|T_{R},T_{R}^{z}\right\rangle $ 

If all the spins in one half of the cluster had equal participation 
in their collective multiplet (observed in the entanglement spectrum),then,
\begin{equation}
\left\langle T_{L},T_{L}^{z}\left| \vec S_{o}\right|T_{L} T_{L}^{'z}\right\rangle = - \left\langle T_{L},T_{L}^{z}\left| \vec S_{e}\right|T_{L} T_{L}^{'z}\right\rangle
\end{equation}
where $e$ ($o$) refers to \emph{any} even(odd) sublattice site in the same half of the bond centered cluster.
Therefore, if one were to create the equally weighted spin operator $\mathbf{T_{L}}=\sum_{i} \vec S_{i}$ 
and consider its matrix elements one would get,
\begin{equation}
\left\langle \vec T_{L} \right\rangle = \pm 2I_{b} \left\langle \vec S_{0} \right\rangle
\label{eq:S_T_relation}
\end{equation}
where 0 refers to the central site in one half of the bond centered cluster. 
The sign depends on whether the central site and the boundary sites are on the same ($+$) 
or opposite ($-$) sublattices.

What happens when the spins are not equally participating in the multiplet?
The simplifying assumption we make here is that within the projected low energy subspace, each individual spin half operator at lattice site $i$, $\vec S_{i}$, is proportional to the operator $ \vec T_{L}/I_{b}$.

Using the fact that  $I_{b} \sim N_{s}$, this relation can be expressed as,
\begin{equation}
\vec S_{0} = \frac{\gamma_{b}}{N_{s}} \mathbf{{T}}_{L} \label{eq:approx_matrix_element_gamma}
\end{equation}
where the constant $\gamma_{b}$ has been used to denote the proportionality factor. 
A similar relation exists for $\vec S_{1}$ and $\vec T_{R}$.

From equation (\ref{eq:approx_matrix_element_gamma}) it follows that,
\begin{equation}
J\left\langle \vec S_{0}\cdot\vec S_{1}\right\rangle = \frac{J\gamma_{b}^{2}}{N_{s}^{2}}\left\langle \mathbf{{T}}_{L} \cdot {\mathbf{T}}_{R}\right\rangle \label{eq:S0_S1_TL_TR_3}
\end{equation}
  
The Hamiltonian $\mathbf{{T}}_{L}\cdot {\mathbf{T}}_{R}$ is
diagonalized by coupling the left and right {}``giant spins'' into
a spin for the whole system i.e. ${\mathbf{T}}={\mathbf{T}}_{L}+{\mathbf{T}}_{R}$,
whose eigenstates are given by $\left|T,T^{z}\right\rangle $.
\begin{eqnarray}
{\left\langle \mathbf{{T}}_{L}\cdot{\mathbf{T}}_{R}\right\rangle}_{\left|T,T^{z}\right\rangle}  & = & \frac{1}{2}\left\langle {\mathbf{T}}^{2}-{\mathbf{T}}_{L}^{2}-{\mathbf{T}}_{R}^{2}\right\rangle \label{eq:diagonalize_TL_dot_TR}\\
 & = & \frac{1}{2}T(T+1)- I_{b}(I_{b}+1)\label{eq:E_T}
\end{eqnarray}
where $T$ varies from $0,1.....,2I_{b}$. Note that $T_{L}$ and $T_{R}$ are constant and equal to $I_{b}$.
The term $I_{b}(I_{b}+1)$ is a harmless constant energy shift to all states. 
Thus, the energy spectrum (up to
a overall shift) as a function of $T$ is simply,
\begin{equation}
E(T)=\frac{J\gamma_{b}^{2}}{2N_{s}^{2}}\; T(T+1)\label{eq:E_vS_T}\\
\end{equation}
This is exactly the Hamiltonian of a quantum rotor with a {}"anomalous"
moment of inertia scaling as $N_{s}^{2}$. This simple picture, hence, rather remarkably explains our numerical
observations in section~\ref{sec:GS_Tower_of_States} of the paper. We find $\gamma_{b}$ to be $\sim 3.24 $ from fits to our numerical data based on finite size scaling of the moment of inertia.

Note that though the giant spins are interacting via a weak bond,
the fact that they are paired up in a singlet ground state makes the state highly entangled. In comparison, as we shall soon see, the ground state of the site-centered tree (in the $S_{z}=S_{0}$ sector)
is closer to a product state of the giant spins and hence has a lower degree of entanglement. (This also explains why the convergence of DMRG calculations is
more rapid with the number of states $M$ for the site-centered case as compared to the bond-centered case.)

\subsubsection{Site-centered tree}
Let us now perform essentially the same analysis for the site centered Cayley tree to 
further shed light on (and validate) the {}"giant spins" picture. 
Rewriting the Heisenberg Hamiltonian we get,
\begin{equation}
H=H_{1}+H_{2}+H_{3}+J\vec S_{0}\cdot\left(\vec S_{1}+\vec S_{2}+\vec S_{3} \right)\label{eq:Heisenberg_Hamiltonian_site_centered}
\end{equation}
where 0 refers to the central site and 1, 2, 3 refer to the
sites connected to it (as has been schematically represented
in Fig.~\ref{fig:giant_spins}). As before, make the substitution of $\vec S_{1}$,$\vec S_{2}$,$\vec S_{3}$ in terms of the giant spins ${\mathbf{T}}_{1}$,${\mathbf{T}}_{2}$,${\mathbf{T}}_{3}$ each of which has spin length $I_{s}$. Then couple the three giant spins into a bigger giant spin ${\mathbf{T}}$. The Hamiltonian now reads as,
\begin{equation}
H=H_{1}+H_{2}+H_{3}+\frac{J\gamma_{s}}{N_{s}}\vec S_{0}\cdot {\mathbf{T}} \label{eq:Heisenberg_Hamiltonian_site_centered_giant}
\end{equation}
where $\gamma_{s}$ is a constant and ${\mathbf{T}}={\mathbf{T}}_{1}+{\mathbf{T}}_{2}+{\mathbf{T}}_{3}$. Note that the angular momentum coupling rules predict that the value of $T$ are in the range from 0,1... to $3I_{s}$. 

Let us now couple the giant spin ${\mathbf{T}}$ to the spin $1/2$ degree of freedom at the center of the cluster. 
The energy (in units of $J\gamma_{s}/N_{s}$ and up to a constant of $-\frac{3}{4}$) for the $S=3I_{s}+\frac{1}{2}$ and $3I_{s}-\frac{1}{2}$ states is given by,
\begin{subequations}
\begin{align}
E^{3I_{s}+\frac{1}{2}}_{3I_{s}}&=&\left(3I_{s}+\frac{1}{2}\right)\left(3I_{s}+\frac{3}{2}\right)-3I_{s}\left(3I_{s}+1\right)\\
E^{3I_{s}-\frac{1}{2}}_{3I_{s}}&=&\left(3I_{s}-\frac{1}{2}\right)\left(3I_{s}+\frac{1}{2}\right)-3I_{s}\left(3I_{s}+1\right)
\end{align}  
\end{subequations}
where the superscript indicates the global value of the spin $S$ and the subscript is used to indicate the $T$ value it was made of.
Similarly we have,
\begin{subequations}
\begin{align}
E^{3I_{s}-\frac{1}{2}}_{3I_{s}-1}&=&\left(3I_{s}-\frac{1}{2}\right)\left(3I_{s}+\frac{1}{2}\right)-3I_{s}\left(3I_{s}-1\right)\\
E^{3I_{s}-\frac{3}{2}}_{3I_{s}-1}&=&\left(3I_{s}-\frac{3}{2}\right)\left(3I_{s}-\frac{1}{2}\right)-3I_{s}\left(3I_{s}-1\right)
\end{align}  
\end{subequations}

Simplifying equations and incorporating the constant of $-\frac{3}{4}$ we get,
\begin{subequations}
\begin{eqnarray}
E^{3I_{s}+\frac{1}{2}}_{3I_{s}}&=&3I_{s}\\
E^{3I_{s}-\frac{1}{2}}_{3I_{s}}&=&-3I_{s}-1\\
E^{3I_{s}-\frac{1}{2}}_{3I_{s}-1}&=&3I_{s}-1\\
E^{3I_{s}-\frac{3}{2}}_{3I_{s}-1}&=&-3I_{s}
\end{eqnarray}  
\end{subequations}

Observe that $E^{3I_{s}-\frac{1}{2}}_{3I_{s}}$ is the lowest energy. This is in concordance with the Lieb-Mattis theorem~\cite{Hulten_Marshall_Lieb_Mattis} i.e. the ground state has total spin $S_{0}=3I_{s}-\frac{1}{2}$. The energy gap (now in absolute units) of the $S_{0}$ to $S_{0}+1$ transition is given by,
\begin{equation}
\Delta\left(S_{0}\rightarrow S_{0}+1 \right)=\frac{J\gamma_{s}}{N_{s}}\left(6I_{s}+1\right)
\end{equation}  

Since $I_{s}$ is approximately $N_{s}/18$ for large $N_{s}$ we get a gap of,
\begin{equation}
\Delta\left(S_{0}\rightarrow S_{0}+1 \right)\approx\frac{J\gamma_{s}}{3}
\label{eq:delta_inf}
\end{equation}

This is consistent with our numerical observation that the gap is finite in the large $N_{s}$ limit. 
Since the measured gap is $\sim 0.73 J$ we infer that $\gamma$ must be $\sim 2.19$.

We give further credibility to our giant spin interpretation by testing the prediction of the gap for the $S_{0}$ to $S_{0}-1$ transition. 
This turns out to be gapless in the large $N_{s}$ limit,
\begin{equation}
\Delta\left(S_{0}\rightarrow S_{0}-1 \right)\approx\frac{J\gamma_{s}}{N_{s}}
\end{equation}
which is consistent with our DMRG calculations 
The measured $\gamma_{s}$ from the fit of the DMRG data to $\Delta=J\gamma_{s}/N_{s}$ is found to be $\sim 2.19$ 
consistent with the estimate from equation (\ref{eq:delta_inf}), serving as another check of the theory.	

\subsubsection{Fibonacci Cayley tree}
The entanglement spectrum of the Fibonacci Cayley tree (see Fig.~\ref{fig:dm_cuts}(c)) indicates the creation of a spin 1/2 degree of freedom
as opposed to the {}"giant spins" encountered previously. The 
cut shown in Fig.~\ref{fig:dm_cuts}(c) shows a region having an imbalance of one, 
which is the maximum possible for any cut. 

We believe the lowest energy excitation of this system involves a breaking of a dimer and creation of two unpaired spins.
Since the bonds in the interior have a decreasing dimerization strength, the energy required to create this excitation
is expected to be vanishingly small in the infinite lattice limit. This is seen in the spin gap in Table~\ref{table:energies},
but an explanation of the observed numerical exponent is a subject of further investigation and beyond the scope of this paper.

\section{Schwinger Boson Mean Field Theory For Singlet Ground states}
\label{sec:SBMFT}
Can we understand the presence or absence of long range order on these trees at the mean field level?
For this we appeal to the Schwinger Boson Mean Field Theory~\cite{AA_PRL} which is capable of 
describing quantum disordered and ordered states within the same 
framework~\cite{SS_Ma,sachdevprb1992}. In this section we will see that this theory 
is a good description of the singlet ground states of the bond centered and Fibonacci trees.
This section also serves to expand the domain of application of the Schwinger Boson formalism 
to situations where multiple parameters need to be optimized simultaneously~\cite{Misguich} 
(such as non uniform systems or systems with very few symmetries).

\subsection{Notation and formal set-up}
\label{sec:SBMFT_notation}
The $SU(N)$ Heisenberg Hamiltonian is expressed in terms of Schwinger
Bosons by defining a bond operator,
\begin{equation}
\mathcal{A}_{ij}=\sum_{m=1}^{N}b_{im}b_{jm}\label{eq:bond operator}
\end{equation}
where each Schwinger boson operator $b_{im}$ carries two labels,
$i$ or $j$ for site indices and $m$ for flavor index. The physical
Heisenberg model equation~\eqref{eq:Heisenberg_Hamiltonian} corresponds to $N=2$. The procedure is to decouple
the quartic bosonic Hamiltonian into a one-body mean field Hamiltonian
by doing an expansion in $1/N$. Solving the mean field Hamiltonian
and putting in $N=2$ allows us to compare the SBMFT results with
DMRG calculations.

The $SU(N)$ Hamiltonian in terms of Schwinger bosons is,
\begin{equation}
\mathcal{H}_{Heis.}=-\frac{J}{N}\sum_{\langle ij\rangle}\left(\mathcal{A}_{ij}^{\dagger}\mathcal{A}_{ij}-2S^{2}\right)\label{eq:SU(N) hamiltonian}
\end{equation}
The mapping of the spin Hamiltonian to Schwinger bosons is exact if
we meet the condition site by site,
\begin{equation}
\sum_{m=1}^{N}b_{im}^{\dagger}b_{im}=NS\label{eq:lambda constraint 0}
\end{equation}
which ensures that the Hilbert space of the bosons is restricted by
the spin size $S$ (and the corresponding $|S,S_{z}\rangle$ states).
However, we will impose this constraint only on the expectation 
$b_{im}^{\dagger}b_{im}\rightarrow\langle b_{im}^{\dagger}b_{im}\rangle_{MF}$.
As a result of not satisfying equation~\eqref{eq:lambda constraint 0} exactly, 
the mean field energy $E_{MF}$ and correlations differ from exact results
(DMRG calculations) by overall scale factors~\cite{AA_PRL,dmonkagome}. 

The decoupled mean field Hamiltonian $\mathcal{H}_{MF}$ is expressed
in terms of the following variational parameters: a set of bond variables
$Q_{ij}$ for every bond $i,j$, Lagrange multipliers $\lambda_{i}$
which impose equation~\eqref{eq:lambda constraint 0} and a condensate field
$\beta_{i}=\delta_{1,m}\langle b_{im}\rangle/\sqrt{N}$ which develops a
non-zero value in a phase with LRO~\cite{sachdevprb1992}. $\mathcal{H}_{MF}$
is then given by~\cite{AAarxiv},
\begin{eqnarray}
\mathcal{H}_{MF} & = & \sum_{i=1}^{N_{s}}\lambda_{i}\left(\sum_{m=1}^{N}b_{im}^{\dagger}b_{im}-NS\right)+\frac{N}{J}\sum_{\langle ij\rangle}|Q_{ij}|^{2}\nonumber \\
 &  & +\sum_{i<j}\left(Q_{ij}\mathcal{A}_{ij}^{\dagger}+Q_{ij}^{*}\mathcal{A}_{ij}\right)\nonumber \\
 &  & +\frac{1}{\sqrt{N}}\sum_{i=1}^{N_{s}}\left(\phi_{i}^{*}b_{i1}+\phi_{i}b_{i1}^{\dagger}\right)\label{eq:mean field hamiltonian}
\end{eqnarray}

The field $\phi_{i}$ couples linearly to Schwinger bosons and is
conjugate to $\delta_{1,m}\langle b_{im}^{\dagger}\rangle$. As a
result of this parametrization, the lowest spinon frequency mode $\omega_{0}$
of the $m=1$ flavor develops a macroscopic occupation of Schwinger
bosons on Bose condensation. At the mean field level the different
boson flavors decouple and the part of the mean field Hamiltonian
quadratic in bosonic operators $b_{im},b_{im}^{\dagger}$ ($m=2,...,N$) can be expressed
as $N-1$ copies of a quadratic Hamiltonian $\mathcal{H}_{MF}^{m}$.
$\mathcal{H}_{MF}^{m}$ is given by,
\begin{equation}
\mathcal{H}_{MF}^{m}=\sum_{i=1}^{N_{s}}\lambda_{i}b_{im}^{\dagger}b_{im}+\sum_{i<j}\left(Q_{ij}b_{im}^{\dagger}b_{jm}^{\dagger}+\mbox{h.c.}\right)\label{eq:hform}
\end{equation}

Integrating out the bosonic fields then gives us a set of
single-spinon frequency modes and the total mean field energy $E_{MF}$. Since we do not have
the luxury of momentum space, we adopt a real space Bogoliubov diagonalization
procedure~\cite{mucciolo}. Since $\mathcal{H}_{MF}^{m}$ is block
diagonal in the flavor basis (the Hamiltonian is the same for all $N-1$ 
flavors) we can now drop the flavor index $m$ and express it as
\begin{equation}
\mathcal{H}_{MF}^{m}=\frac{1}{2}\left(\begin{array}{cc}
\mathbf{b}^{\dagger} & \mathbf{b}\end{array}\right)\left(\begin{array}{cc}
\boldsymbol{\Lambda} & \boldsymbol{Q}\\
\boldsymbol{Q} & \boldsymbol{\Lambda}
\end{array}\right)\left(\begin{array}{c}
\mathbf{b}\\
\mathbf{b}^{\dagger}
\end{array}\right)\label{eq:matrixformofhm}
\end{equation}
where $m\neq1$, $\mathbf{b}=\left(b_{1},b_{2},...,b_{N_{s}}\right)$ and $\boldsymbol{\Lambda}$
and $\boldsymbol{Q}$ are $N_{s}\times N_{s}$ matrices given by $\boldsymbol{\Lambda}_{ij}=\lambda_{i}\delta_{ij}$
and $\boldsymbol{Q}_{ij}=Q_{ij}$ for nearest neighbor sites $ i,j$. $\mathcal{H}_{MF}^{m}$
can now be diagonalized by introducing Bogoliubov transfomation matrices,
$ $$\mathbf{U}$ and $\mathbf{V}$ defined as follows,
\begin{equation}
\left(\begin{array}{c}
\mathbf{b}\\
\mathbf{b^{\dagger}}
\end{array}\right)=\left(\begin{array}{cc}
\mathbf{U} & \mathbf{V}\\
\mathbf{V}^{*} & \mathbf{U}^{*}
\end{array}\right)\left(\begin{array}{c}
\boldsymbol{\alpha}\\
\boldsymbol{\alpha}^{\dagger}
\end{array}\right)\label{eq:bogoliubovtransformation}
\end{equation}
where $\boldsymbol{\alpha}=\left(\alpha_{1},\alpha_{2},...,\alpha_{N_{s}}\right)$
is a vector of Bogoliubov quasiparticle annihilation operators. Each quasiparticle creation (annihilation) operator $\alpha_{\mu}^{\dagger}$ ($\alpha_{\mu}$) creates (destroys) a bosonic quasiparticle in the single particle mode $\mu$, where $\mu$ goes from $1$ to $N_{s}$. The transformation (\ref{eq:bogoliubovtransformation})
allows us to switch to the Bogoliubov quasiparticle basis where $\mathcal{H}_{MF}^{m}$
becomes, 
\begin{equation}
\mathcal{H}_{MF}^{m}=\sum_{\mu=1}^{N_{s}}\omega_{\mu}\left(\alpha_{\mu}^{\dagger}\alpha_{\mu}+\frac{1}{2}\right)\label{eq:diagonalizedhmf}
\end{equation}
We can now perform a Legendre transformation to replace the field $\phi_{i}$ by $\beta_{i}$, allowing us to express $E_{MF}$ as, 
\begin{equation}
\frac{E_{MF}}{N}=\sum_{\mu=1}^{N_{s}}\frac{1}{2}\omega_{\mu_{i}}+\frac{1}{J}\sum_{i>j}|Q_{ij}|^{2}-\left(S+\frac{1}{2}\right)\sum_{i=1}^{N_{s}}\lambda_{i}+E_{cond}\label{eq:emf}
\end{equation}
and $E_{cond}$ is given by,
\begin{equation}
E_{cond}=\sum_{i=1}^{N_{s}}\lambda_{i}|\beta_{i}|^{2}+\sum_{i<j}\left(Q_{ij}\beta_{i}^{*}\beta_{j}^{*}+\mbox{h.c.}\right)\label{eq:econd}
\end{equation}
We now consider the case of $N=2$. The variational parameters $\{\lambda_{i}\},\{Q_{ij}\}$ and $\{\beta_{i}\}$
are determined by minimizing $E_{MF}$ with respect to each of them
giving the following constraints,
\begin{subequations}
\begin{eqnarray}
\frac{\partial E_{MF}}{\partial\lambda_{i}}=0\Rightarrow|\beta_{i}|^{2}+\langle b_{i2}^{\dagger}b_{i2}\rangle=S\label{eq:lambda constraint}\\
\frac{\partial E_{MF}}{\partial Q_{ij}}=0\Rightarrow\beta_{i}^{*}\beta_{j}^{*}+\frac{\langle \mathcal{A}_{ij}\rangle}{N}+\frac{1}{J}Q_{ij}=0\label{eq:qij constraint}\\
\frac{\partial E_{MF}}{\partial\beta_{i}}=0\Rightarrow\lambda_{i}\beta_{i}^{*}+\sum_{j\mbox{ n.n.}i}Q_{ij}^{*}\beta_{j}^{*}=0\label{eq:beta constraint}\\
\nonumber
\end{eqnarray}
\end{subequations}
One of the obvious considerations of applying this theory to such
a non-uniform lattice is the large number of variational parameters
which, in general, scale with the system size $N_{s}$. However, due
to the symmetries of the Cayley tree the number of independent parameters are reduced to order $g$.
The task then is to find an optimal set of parameters $\{\lambda_{i}^{*},Q_{ij}^{*},\beta_{i}^{*}\}$
which satisfy the constraints in (\ref{eq:lambda constraint}), (\ref{eq:qij constraint})
and (\ref{eq:beta constraint}). This is done numerically and is
discussed in subsection C. The resulting optimal $E_{MF}(\lambda_{i}^{*},Q_{ij}^{*},\beta_{i}^{*})$
can be related to the physical Heisenberg energy via,
\begin{equation}
E_{Heis.}=2E_{MF}+\sum_{\langle ij\rangle}JS^{2}\label{eq:eheis}
\end{equation}
A note on the $\beta_{i}$ minimization constraint (\ref{eq:beta constraint}):
a trivial solution of this equation is to choose $\beta_{i}=0$ for
all sites. This is the quantum disordered phase. A non zero value
of $\beta_{i}$ signals condensation of Schwinger bosons and long
range order. 

For finite uniform systems there is no spontaneous symmetry breaking and correspondingly no condensation of bosons~\cite{Auerbach}. 
Introduction of the condensate field $\beta_{i}$ is analogous to applying a staggered field 
in the system that couples to the staggered magnetization. This breaks the degeneracy of the single particle energies
corresponding to the two boson flavors (for $N=2$). The condensate fraction begins to build up in the flavor mode with the lowest frequency.

The algorithm tries to initially find a self-consistent mean field solution by varying only the set of $\lambda_{i}$ and $Q_{ij}$'s.
However, if we cannot satisfy the constraints in equations (\ref{eq:lambda constraint}, \ref{eq:qij constraint}) (with $\beta_{i}=0$), 
we resort to adding the condensate field $\beta_{i}$, as an additional set of variational parameters. While we can not completely rule out the 
possibility of a solution with $\beta_{i}=0$, 
we believe that the appearance of a condensate is \emph{physical}. 

\subsection{Correlation functions in Quantum disordered and LRO phase}
\label{sec:SBMFT_Corr}
Here we compute correlation functions that enter into
the self consistency equations (\ref{eq:lambda constraint}, \ref{eq:qij constraint}, \ref{eq:beta constraint}). 
The boson density of a given flavor at site $i$ is given by (suppressing the flavor index),
\begin{eqnarray}
\langle b_{i}^{\dagger}b_{i}\rangle & = & \sum_{p,q=1}^{N_{s}}\langle\left(\mathbf{U}_{ip}^{*}\boldsymbol{\alpha}_{p}^{\dagger}+\mathbf{V}_{ip}^{*}\boldsymbol{\alpha}_{p}\right)\left(\mathbf{U}_{iq}\boldsymbol{\alpha}_{q}+\mathbf{V}_{iq}\boldsymbol{\alpha}_{q}^{\dagger}\right)\rangle\nonumber \\
 & = & \sum_{p,q=1}^{N_{s}}\mathbf{V}_{ip}^{*}\mathbf{V}_{iq}\langle\boldsymbol{\alpha}_{p}\boldsymbol{\alpha}_{q}^{\dagger}\rangle\nonumber \\
 & = & \left(\mathbf{V}^{*}\mathbf{V}^{T}\right)_{ii}\label{eq:bosondensity}
\end{eqnarray}
The indices $p,q$ run over all single particle modes and we made use of $\langle\boldsymbol{\alpha}_{p}\boldsymbol{\alpha}_{q}^{\dagger}\rangle=\delta_{p,q}$
which follows since the SBMFT many body ground state is the vacuum of Bogoliubov quasiparticles.
Similarly, 
\begin{equation}
\langle b_{i}b_{j}\rangle=\left(\mathbf{U}\mathbf{V}^{T}\right)_{ij}\label{eq:expectationogbibj}
\end{equation}
Spin correlation functions $G_{ij}$ can be computed in a similar
fashion. The only complication arises in the $SU(2)$ broken symmetry
phase where, due to loss of spin rotational invariance, we need to
compute the $\langle S_{i}^{z}S_{j}^{z}\rangle$ correlations. This
involves evaluating a quartic expectation which we decouple into a
series of 2 point functions using Wick's theorem. For simplicity of
notation we define the following combinations of $\mathbf{U}$ and
$\mathbf{V}$ matrices,
\begin{eqnarray}
\tilde{\mathbf{Q}} & \equiv & \mathbf{U}\mathbf{V}^{T}\nonumber \\
\tilde{\mathbf{U}} & \equiv & \mathbf{U}\mathbf{U}^{\dagger}\nonumber \\
\tilde{\mathbf{V}} & \equiv & \mathbf{V}\mathbf{V}^{T}\label{eq:matrixdef}
\end{eqnarray}
Spin correlations in the Quantum disordered and the broken symmetry LRO
phase are then given by,
\begin{subequations}
\begin{eqnarray}
\left(G_{ij}\right)_{Q.dis.} & = & \left\{ \begin{array}{c}
-\frac{3}{2}\tilde{\mathbf{Q}}_{ij}^{\dagger}\tilde{\mathbf{Q}}_{ij}\\ \:\mbox{ for \ensuremath{i\in}}A,j\in B\\ \\
\frac{3}{2}\tilde{\mathbf{V}}_{ij}\tilde{\mathbf{U}}_{ij}\\\ \mbox{ for }(i,j)\in A \mbox{ or } B\\ \\
\end{array}\right.\\
\left(G_{ij}\right)_{LRO} & = & \left\{ \begin{array}{c}
-N\beta_{i}\beta_{j}\tilde{\mathbf{Q}}_{ij}-\frac{1}{4}\tilde{\mathbf{Q}}_{ij}^{\dagger}\tilde{\mathbf{Q}}_{ij}\\ \:\mbox{for \ensuremath{i\in}}A,j\in B\\ \\
\frac{1}{2}N\beta_{i}\beta_{j}\left(\tilde{\mathbf{U}}+\tilde{\mathbf{V}}\right)_{ij}+\frac{1}{4}\tilde{\mathbf{V}}_{ij}\tilde{\mathbf{U}}_{ij}\\ \:\mbox{ for } (i,j)\in A \mbox{ or } B\\ \\
\end{array}\right.\\
\nonumber
\label{eq:spincorrelations}
\end{eqnarray}
\end{subequations}
In the quantum disordered phase SBMFT overestimates~\textbf{ $G_{ij}$}
by an overall scale factor~\cite{AA_PRL} of $3/2$. We normalize the SBMFT correlation
function by this factor to take into account the $1/N$ fluctuation
effects for $N=2$. Similarly, in the phase with LRO we find that
we need to scale up $G_{ij}$ by a factor of $1.8$ to make it agree
quantitatively with DMRG results. Similar overall scale factors have been reported previously~\cite{dmonkagome}. These overall scale factors can be suppressed by using Gutzwiller projected 
mean field wavefunctions~\cite{Tay}, which is feasible only for small system sizes.
Such projected wavefunctions have also been shown to give energies in agreement with exact results~\cite{Tay,dmonkagome}. 

\subsection{Numerical Implementation}
Using the symmetry of the bond-centered Cayley tree, we reduce the number
of variational mean field parameters. A first simplification results
from the fact that all sites within a given shell on the lattice are
equivalent and are therefore assigned the same $\lambda_{i}$ and
$\beta_{i}$. Similarly, all bonds connecting two successive shells
are equivalent and have the same $Q_{ij}$'s. For the Fibonacci cluster 
there are fewer exact symmetries (only reflection about the central bond) 
compared to the Cayley tree and therefore a larger number of variational parameters are required. 

We use the Nelder Mead simplex optimizer from the GSL library to minimize the following weighted combined cost function 
which aims to reduce (\ref{eq:emf}) subject to the constraints (\ref{eq:lambda constraint}), (\ref{eq:qij constraint}) and (\ref{eq:beta constraint}). 
Since each of these constraint equations are obtained by minimizing (\ref{eq:emf}) 
with respect to the variational parameters $\lambda_{i}$, $Q_{ij}$ and $\beta_{i}$, enforcing the constraints will minimize $E_{MF}$.
\begin{equation}
C\left(\{\lambda_{i}\},\{Q_{ij}\},\{\beta_{i}\}\right)=\mu_{0}C_{\lambda}+\mu_{1}C_{Q}+\mu_{2}C_{\beta}\label{eq:cost function}
\end{equation}
where $\{\mu_{0},\mu_{1},\mu_{2}\}$ are relative weights of terms
in the cost function and the costs are given by,
\begin{subequations}
\begin{eqnarray}
C_{\lambda} &=&\sum_{i=1}^{N_{s}}\left(\left(|\beta_{i}|^{2}+\langle b_{i2}^{\dagger}b_{i2}\rangle\right)-S\right)^{2}\label{eq:lambdaincostfunc}\\
C_{Q}&=&\sum_{i<j}\left(\beta_{i}\beta_{j}+\frac{1}{N}\langle\mathcal{A}_{ij}\rangle+\frac{1}{J}Q_{ij}\right)^{2}\label{eq:qcostincostfunc}\\
C_{\beta}&=&\sum_{i=1}^{N_{s}}\left(\lambda_{i}\beta_{i}^{*}+\sum_{j\mbox{ n.n.}i}Q_{ij}^{*}\beta_{j}^{*}\right)^{2}\label{eq:betacostincostfunc}
\end{eqnarray}
\end{subequations}

In practice, to minimize the weighted cost function (\ref{eq:cost function}),
tolerance values for the $C_{\beta}$ and $C_{\lambda}$ are set at $10^{-8}$ and $10^{-14}$ respectively 
and the $Q_{ij}$'s are solved for self-consistently.
A good initial guess for the $Q_{ij}$'s is a pattern that favors dimerization as suggested by results from Exact Diagonalization for small clusters. 
A good rule of thumb for the bond centered cluster is to begin by dimerizing (assigning a high value of $Q_{ij}$) 
the outermost bond and to create a pattern, moving inwards, of alternating bond strengths.

For cases requiring a larger number of variational parameters (like in the case of the Fibonacci cluster) 
it helps to guide the Nelder Mead optimization using a "relaxation" algorithm. 
The algorithm starts with an initial guess for the  $Q_{ij}$'s and allows the optimizer to find an optimal set of $\lambda_{i}$'s. 
If the tolerance level for $C_{\lambda}$ is not met, the  $Q_{ij}$'s are allowed to relax: 
i.e. the best set of $\lambda_{i}$'s is taken as initial guess in an optimization where 
the  $Q_{ij}$'s are now taken to be the variational parameters. 
Thus, by alternating in the choice of variational parameters between the $\lambda_{i}$ and the $Q_{ij}$ for each optimization cycle we converge to the desired tolerance limit. 

The stability of the obtained mean field solution was checked by adding small random perturbations 
to the optimal mean field parameters. 
In the quantum disordered phase, the saddle point was checked to 
correspond to a maximum in the $\lambda_{i}$ and a minimum in the $Q_{ij}$.

Every optimization cycle scales as $\sim N_{s}^{3} \tau$, where $\tau$ is the total time for 
functional evaluation taken by the optimizer to converge to a solution.
A typical optimization time for the $N_{s}$=126 cluster is about 20 minutes on a personal workstation (2.7 GHz Intel Core i7).

\subsection{Results}
\label{sec:SBMFT_Results}
The landscape of the cost function (\ref{eq:cost function}) 
in parameter space has many local
minima with very similar mean field energies (differing by $1\%$ or less).
As a result, we get a zoo of physically plausible mean field solutions,
all of which satisfy (\ref{eq:lambda constraint}) and have comparable
$E_{MF}.$ To choose the optimal solution from amongst those, we look
at the $\beta_{i}$ minimization constraint (\ref{eq:beta constraint})
and hand pick the one which has the lowest $C_{\beta}$. In other
words, the chosen solution has the lowest spinon frequency $\omega_{0}$. 

The mean field energy and correlation functions for the bond-centered Cayley trees
suffer from significant finite size effects (number of sites on the
boundary scales as the system size $N_{s}$). As a result, for the
finite systems considered above, the lowest spinon frequency is always gapped 
$\omega_{0}\neq0$ in spite of a very low $\beta_{cost}(\sim10^{-8})$. 
However, with increasing system size, $\omega_{0}$
lowers and spinons begin to condense in this lowest frequency mode.
A good (to within $5\%$) fit to the lowest spinon frequency versus system size $N_{s}$ plot is given by the function $\omega_{0}(N_{s})=0.067/N_{s}+1.626/ N_{s}^{2}$. Spinon condensation and a very small $\omega_{0}$ suggest long range order in the thermodynamic limit.
\begin{figure}[htpb]
\includegraphics[width=0.95\linewidth]{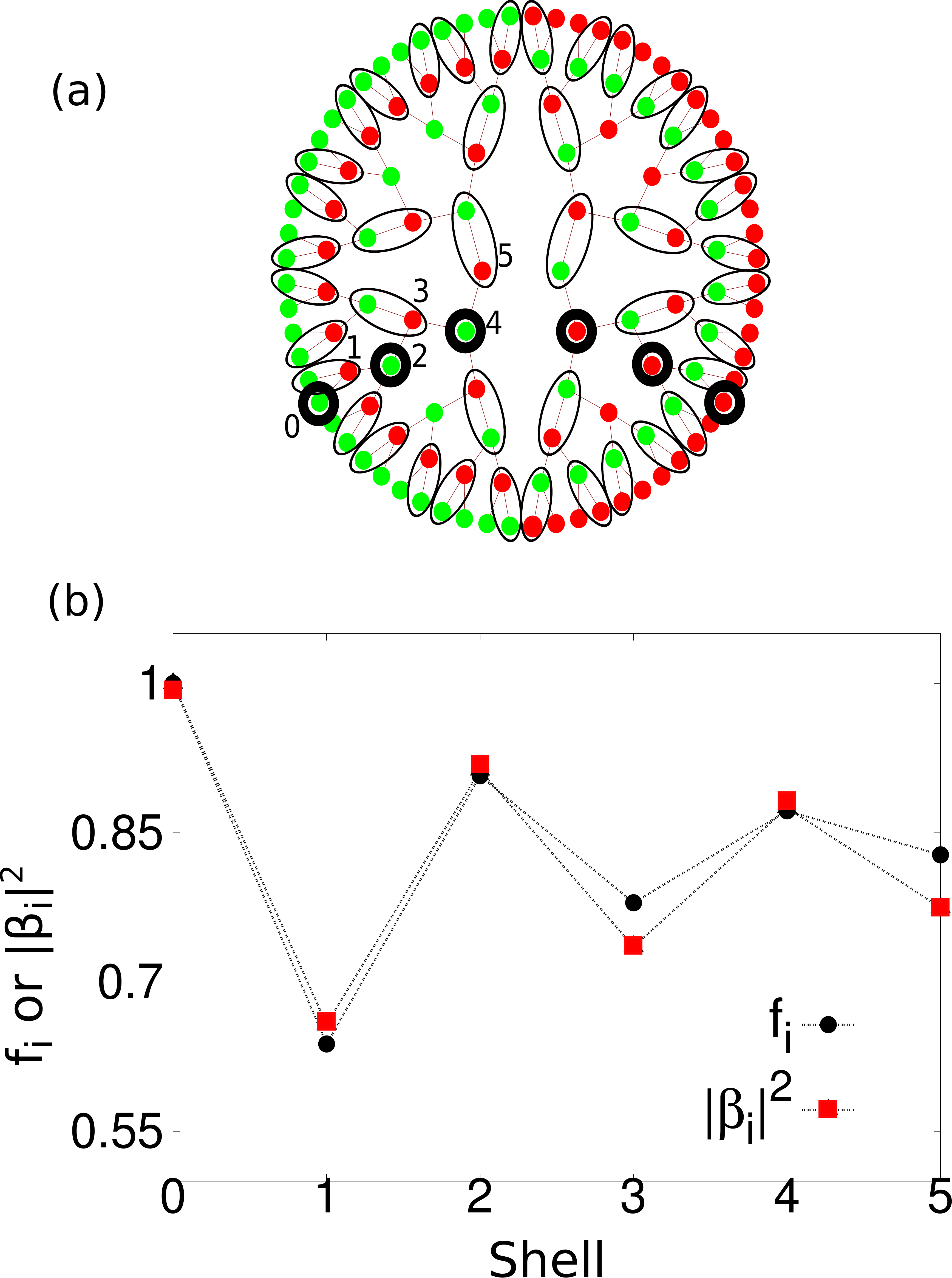}
\caption{(Color online) Above (a): Heuristic for computing the number of {}"dangling spins"
as proposed by Wang and Sandvik. Below (b): The magnitude of the {}"flippability" as in equation~\eqref{eq:c_i_defn} 
computed from DMRG and the condensate fraction $|\beta_{i}|^{2}$ 
computed from SBMFT on every shell of the bond centered Cayley tree. 
Both quantities are qualitatively consistent with each other and confirm 
the {}"dangling" spin heuristic shown above.}
\label{fig:dangling}
\end{figure}

Since condensation of spinons signals long range order, 
sites with a higher condensate fraction have a greater participation in establishing Neel order 
on the cluster. By mapping out the condensate fraction on 
different radial shells in Fig.~\ref{fig:dangling}(b) we notice the strong correspondence between sites 
with large condensate densities and those with a high {}"flippability" as in equation~\eqref{eq:c_i_defn}.

Our results can be put in perspective with respect to a heuristic for computing the number of {}"dangling spins"
proposed by Wang and Sandvik~\cite{Wang_Sandvik_1} who encountered the role of sublattice imbalance in the context of percolation clusters on the square lattice. 
In their picture, geometric constraints of a diluted square lattice forbid spins from pairing up with their neighbors into dimers, 
leaving some of them unpaired or {}"dangling"~\cite{Wang_Sandvik_1}. They believed these emergent spins to be the effective low energy degrees of freedom. 
In a similar spirit we {}``dimerize" the lattice maximally 
as shown in Fig.~\ref{fig:dangling}(a) and
the spins that remain are called {}"dangling". These are the representative
spins participating as the low energy degrees of freedom.
(Note that the choice of maximal dimer covering
is not unique but the number of uncovered spins is the same in all
such coverings) 

\section{Conclusion}
\label{sec:Conclusion}
In this paper, we have explored the relationship between sublattice imbalance and 
nature of the low energy spectrum of 
the bond-centered, site-centered and Fibonacci Cayley 
trees. 

For the bond-centered Cayley tree, we 
find that the spin gap scales with size as $1/N_{s}^{\alpha}$ where $\alpha$ was found to be $\approx$ 2.
We discover an entire tower of states (Fig.~\ref{fig:rotorspectrum}) 
with a much larger moment of inertia (sec.~\ref{sec:GS_Tower_of_States}) than the Anderson Tower of States. 
This low energy scale persists up to a spin value of $S^{*}=2 I_{b}$, where $I_{b}$ refers to a measure of the imbalance
(or the number of {}"unpaired spins") on the bond centered tree (as in equation~\eqref{eq:Imbalance_bond_centered}).

To highlight the role of sublattice imbalance, we introduced the Fibonacci Cayley tree in sec.~\ref{sec:Model}, 
which does not have any locally unpaired spins.
We found it lacks the low lying states characteristic 
of the bond-centered tree (see Fig.~\ref{fig:rotorspectrum}). Instead, the spin gap vanishes as $\approx 1/N_{s}^{0.6}$.
However, both trees have similar susceptibilities ($\sim N_{s}$) at sufficiently large magnetic fields.
This is because the strength of the dimerization is relatively weak at sufficiently high energy scales (comparable to $J$), allowing 
all spins to lock together, leading to an extensive susceptibility.
 
For the site centered tree, our results are 
in good agreement with a recent study~\cite{Kumar_Ramasesha_Soos}.
We report a finite spin gap of $\Delta=0.731(4)$ in the infinite lattice 
limit and a ground state energy of $e_{0}=-0.393854(1)$.

Our results can be explained within a unifying framework of individual spins coupling together to form
collective spin degrees of freedom which we refer to as {}"giant spins". 
The idea for coupling big sublattice spins is well known~\cite{Lhuillier}
in the context of regular lattices. However, we emphasize that the "giant spins" are created by coupling all spins 
(both even and odd sublattice spins) in regions with large local sublattice imbalances. 
For the bond- and site-centered lattices, we find that all lattice sites have a 
(nearly) uniform participation in the "giant spins".
This picture is developed in section \ref{sec:SMA} using the single mode approximation for the excited (triplet) state and 
observing the entanglement spectrum. 

In a broader context, our study aims to understand the nature of unpaired spins coupling 
in a strongly dimerized background. Such spins, termed as "dangling spins", have been predicted 
from numerical simulations~\cite{Wang_Sandvik_1} 
for systems with quenched random dilution.
Thus, a natural extension of our
study is to consider such dangling spins on percolation clusters
where we expect that they couple hierarchically to form emergent giant
spins at different length scales. 
This will be the subject of a future publication~\cite{Changlani}. 

In this paper, we have explored several techniques to develop our understanding. To begin with, we 
obtain accurate many-body wavefunctions and their expectations using an implementation of
the Density Matrix Renormalization Group (DMRG) procedure (described in section~\ref{sec:DMRG})
that works for generic trees. In general, this procedure is expected to be 
well suited to hierarchical lattices and those with few or no loops, such as 
the Husimi Cactus~\cite{Husimi_SL,Cactus}, a lattice of corner sharing triangles (whose centers form the vertices of the Cayley tree), 
and which is locally like the kagome lattice).

To have an understanding at the mean field level, we adapted 
Schwinger Boson Mean Field Theory (SBMFT) 
to a larger number of variational parameters than considered previously~\cite{Tay,dmonkagome,messio}. 
We were able to study spin correlations of the bond-centered and Fibonacci trees
(singlet ground states). Rather remarkably, the theory is quite accurate \emph{quantitatively} in predicting ground state
spin-spin correlation functions (see Fig.~\ref{fig:sample_corrs}), up to overall multiplicative scale factors (as discussed in section~\ref{sec:SBMFT}).
The recipe outlined in
section~\ref{sec:SBMFT} can be used to navigate through the zoo of feasible mean field
solutions by giving relative weights to the constraint equations (\ref{eq:lambdaincostfunc}, \ref{eq:qcostincostfunc}, \ref{eq:betacostincostfunc}).

We believe that most applications of SBMFT have
focused on quantum disordered phases~\cite{Wen,Fa}, but the broken symmetry phase has received less attention. 
The setup can also be generalized to handle frustrated spin systems 
without the need to have an ansatz for the mean field 
fluxes or the decoupling parameters~\cite{Wen}. This might lead to novel 
spin liquid ground states with new kinds
of broken symmetries~\cite{Misguich}. 

After writing this paper,
we came across recent results by G. Misguich~\cite{Misguich}, who has done an extensive numerical study
of SBMFT formalism for spatially inhomogenous states, mostly concentrating on gapped systems. 
However, his study differs from ours in that we include the condensate field as a variational parameter. 
It will be interesting to apply our formalism to further investigate his proposed set of ground states.

\section{Acknowledgment}
\label{sec:Ack}
HJC would like to thank G.K. Chan and S. Sharma
for discussions on the DMRG technique, N. Nakatani for numerical
checks against his tree tensor network code and C.J. Umrigar and
the Cornell Physics Department for use of their computational facilities.
We thank S. Pujari, L. Wang, S. Parameswaran, A. Sandvik, A. Auerbach, D. Arovas and M. Lawler 
for discussions and J. Chalker for his suggestion to study the bond centered Cayley tree. 
We also thank R. Lamberty for 
critically reviewing the manuscript.
HJC, SG and CLH acknowledge support from National Science Foundation grant NSF DMR-1005466. This work made use of the Cornell Center for Materials Research computer facilities 
which are supported through the NSF MRSEC program (DMR-1120296).

\appendix
\renewcommand{\theequation}{{A}.\arabic{equation}}
\section{Derivation of the SMA gap equation for the Heisenberg Model}
\label{sec:derivSMA}
In section~\ref{sec:SMA}, we introduced the single mode approximation (SMA) for the excited state in terms of the ground state,
\begin{equation}
\left | 1' \right\rangle=\sum_{i}u_{i}S^{+}_{i}\left|0\right\rangle
\label{eq:SMA_ansatz_revise}
\end{equation}
where $\left |0 \right\rangle$ is the singlet ground state and $\left| 1' \right\rangle$ is 
the approximate triplet excited state. $i$ refers to a site index.
In this Appendix, we will derive an expression for the SMA energy gap $\Delta_{SMA}$ in terms of the ground state correlations 
and the parameters $u_{i}$.

The expression for the gap between the ground and first excited state is, 
\begin{equation}
\Delta=\frac{\left\langle 1'\left|H\right|1'\right\rangle }{\left\langle 1'\left|1'\right.\right\rangle }-E_{0}\label{eq:SMA_gap_appendix}
\end{equation}
where $E_{0}$ is the ground state energy. Plugging in the expression
for $\left|1'\right\rangle$ from equation (\ref{eq:SMA_ansatz_revise}) we have,
\begin{eqnarray}
\Delta_{SMA} & = & \frac{\left\langle 0\left|\sum_{ij}u_{i}u_{j}S_{j}^{-}HS_{i}^{+}\right|0\right\rangle }{\left\langle 0\left|\sum_{ij}u_{i}u_{j}S_{j}^{-}S_{i}^{+}\right|0\right\rangle }-E_{0}\label{eq:sma_gap_subst}\\
 & = & \frac{\sum_{ij}u_{i}u_{j}\left\langle S_{j}^{-}\left[H,S_{i}^{+}\right]\right\rangle }{\sum_{ij}u_{i}u_{j}\left\langle S_{j}^{-}S_{i}^{+}\right\rangle }\nonumber \\
 &  & +\frac{\sum_{ij}u_{i}u_{j}\left\langle S_{j}^{-}S_{i}^{+}H\right\rangle }{\sum_{ij}u_{i}u_{j}\left\langle S_{j}^{-}S_{i}^{+}\right\rangle }-E_{0}\\
 & = & \frac{\sum_{ij}u_{i}u_{j}\left\langle S_{j}^{-}\left[H,S_{i}^{+}\right]\right\rangle }{2\sum_{ij}u_{i}u_{j}\left\langle S_{j}^{z}S_{i}^{z}\right\rangle }\label{eq:sma_gap_commutator}
\end{eqnarray}
 
Next, consider the nearest neighbor Heisenberg Hamiltonian,
\begin{eqnarray}
H & = & \frac{J}{2}\sum_{\left\langle k,l \right\rangle}\left(S_{k}^{z}S_{l}^{z}+\frac{1}{2}\left(S_{k}^{+}S_{l}^{-}+S_{k}^{-}S_{l}^{+}\right)\right)\label{eq:Heisenberg_model_spluS_sminus}
\end{eqnarray}
where $\left\langle k,l \right\rangle$ refer to nearest neighbor pairs. We have included a
factor of 1/2 outside to compensate for counting each nearest neighbor
term twice. 

We now calculate $[H,S_{i}^{+}]$ occurring in equation~\eqref{eq:sma_gap_commutator}). To do so, we calculate $\left[\vec S_{k}\cdot \vec S_{l},S_{i}^{+}\right]$ as,
\begin{eqnarray}
\left[\vec S_{k}\cdot \vec S_{l},S_{i}^{+}\right] & = & S_{k}^{z}\left[S_{l}^{z},S_{i}^{+}\right]+S_{l}^{z}\left[S_{k}^{z},S_{i}^{+}\right]+\nonumber \\
 &  & \frac{1}{2}S_{k}^{+}\left[S_{l}^{-},S_{i}^{+}\right]+\frac{1}{2}\left[S_{k}^{-},S_{i}^{+}\right]S_{l}^{+}\nonumber \\
 & = & \delta_{il}S_{k}^{z}S_{l}^{+}+\delta_{ik}S_{l}^{z}S_{k}^{+}\nonumber \\
 &  & -\delta_{il}S_{k}^{+}S_{l}^{z}-\delta_{ik}S_{k}^{z}S_{l}^{+}
\label{eq:simple_comm}
\end{eqnarray}

The numerator of equation~\eqref{eq:sma_gap_commutator} involves the term $\left\langle S_{j}^{-}\left[H,S_{i}^{+}\right]\right\rangle$. 
Hence we now consider the action of the $S_{j}^{-}$ operator on the simplified expression 
for $\left[\vec S_{k}\cdot \vec S_{l},S_{i}^{+}\right]$ in equation~\eqref{eq:simple_comm}.
Consider only the terms that have $j=k$ or $j=l$ (since $k = l$ terms
do not occur in the Hamiltonian we do not have to worry about the
possibility $j=k=l$). In addition, time reversal symmetry of the ground state wavefunction (equivalent to simply asserting the $S_{z}$
$\rightarrow$ $-S_{z}$ symmetry of the ground state) ensures that if both $j\neq k$ and $j \neq l$ 
then the three point correlation function is exactly 0. 
This latter point is rather subtle and so we expand on this in Appendix B.

Thus the expression for $S_{j}^{-}\left[\vec S_{k}\cdot \vec S_{l},S_{i}^{+}\right]$ (after retaining only the $j=k$ and $j=l$ terms) is,
\begin{align}
S_{j}^{-}\left[\vec S_{k}\cdot \vec S_{l},S_{i}^{+}\right]=-\delta_{ik}\delta_{jl}\left(\frac{1}{2}-S_{l}^{z}\right)S_{k}^{z}-\delta_{ik}\delta_{jk}\frac{S_{k}^{-}S_{l}^{+}}{2}\nonumber \\
-\delta_{il}\delta_{jk}\left(\frac{1}{2}-S_{k}^{z}\right)S_{l}^{z}-\delta_{il}\delta_{jl}\frac{S_{l}^{-}S_{k}^{+}}{2}\nonumber \\
+\delta_{jk}\delta_{il}\frac{S_{k}^{-}S_{l}^{+}}{2}+\delta_{jl}\delta_{il}\left(\frac{1}{2}-S_{l}^{z}\right)S_{k}^{z}\nonumber \\
+\delta_{jk}\delta_{ik}\left(\frac{1}{2}-S_{k}^{z}\right)S_{l}^{z}+\frac{1}{2}S_{k}^{+}S_{l}^{-}\delta_{ik}\delta_{jl}\label{eq:simplify_correlation}
\end{align}

Inserting (\ref{eq:simplify_correlation}), in the expression for the SMA gap (\ref{eq:sma_gap_commutator}) for the Heisenberg Hamiltonian and utilizing $\left\langle S_{i}^{z}\right\rangle=0$ for all $i$, we obtain,
\begin{equation}
\Delta_{SMA}=-\frac{\sum_{\left\langle k,l \right\rangle}(u_{k}-u_{l})^{2}\left\langle S_{k}^{z}S_{l}^{z}\right\rangle }{2\sum_{ij}u_{i}u_{j}\left\langle S_{i}^{z}S_{j}^{z}\right\rangle }\label{eq:SMA_final}
\end{equation}

\renewcommand{\theequation}{{B}.\arabic{equation}}
\label{sec:proofsss}
\section{Why is $\left\langle \psi\right|S_{j}^{-}S_{k}^{+}S_{l}^{z}\left|\psi\right\rangle=0$ for distinct $j,k,l$ ?}
To derive the SMA gap equation in Appendix~\ref{sec:derivSMA}, we used,
\begin{equation}
\left\langle \psi\right|S_{j}^{-}S_{k}^{+}S_{l}^{z}\left|\psi\right\rangle=0
\end{equation}
for distinct site indices $j,k,l$.
In this Appendix, we will prove this statement for \emph{any} wavefunction which is invariant under time reversal.

Consider three distinct spins $i,j,k$. Express the wavefunction in the basis spanned by the three spins at sites $j,k,l$ and the rest of the spins (collectively termed as {}"environment" $e$),
\begin{equation}
\left|\psi\right\rangle =\sum_{s'_{j}s'_{k}s'_{l}}\sum_{e}w_{e}^{s'_{j}s'_{k}s'_{l}}\left|s'_{j}s'_{k}s'_{l}\right\rangle \otimes\left|e\right\rangle 
\label{eq:wf_jkl}
\end{equation}
Since this wavefunction is an eigenstate of the Heisenberg model (with no external magnetic fields), it follows that under 
time reversal (denoted by operator $T$) we have,
\begin{equation}
\psi \rightarrow z\psi
\label{eq:psi_mpsi}
\end{equation}
where $z$ is $\pm 1$.

This implies that the coefficients in the wavefunction satisfy the relation,
\begin{equation}
w_{e}^{s'_{j}s'_{k}s'_{l}}= z w_{-e}^{-s'_{j}-s'_{k}-s'_{l}} 
\label{eq:w_relation}
\end{equation}

The action of the operator $S_{j}^{-}S_{k}^{+}S_{l}^{z}$ on $\left|\psi\right\rangle$ from equation~\eqref{eq:wf_jkl} yields,
\begin{equation}
S_{j}^{-}S_{k}^{+}S_{l}^{z}\left|\psi\right\rangle =\sum_{s'_{l}}\sum_{e}w_{e}^{\uparrow\downarrow s'_{l}}S_{l}^{'}\left|\downarrow\uparrow s'_{l}\right\rangle \otimes\left|e\right\rangle 
\label{eq:sjkl_wf_jkl}
\end{equation}

Now acting equation~\eqref{eq:sjkl_wf_jkl} with $\left\langle \psi\right|$ from the left and using the orthogonality of the basis we get,
\begin{eqnarray}
\left\langle \psi\right|S_{j}^{-}S_{k}^{+}S_{l}^{z}\left|\psi\right\rangle  & = & \sum_{s'_{l}}\sum_{e}w_{e}^{\downarrow\uparrow s'_{l}}w_{e}^{\uparrow\downarrow s'_{l}}S_{l}^{'}\\
 & = & \frac{1}{2}\sum_{e_{-}}w_{e_{-}}^{\downarrow\uparrow\uparrow}w_{e_{-}}^{\uparrow\downarrow\uparrow} \nonumber \\ & -&\frac{1}{2}\sum_{e_{+}}w_{e_{+}}^{\downarrow\uparrow\downarrow}w_{e_{+}}^{\uparrow\downarrow\downarrow}\label{eq:psi_S_psi}
\end{eqnarray}
where $e_{+}(e_{-})$ reflects the fact that the environment carries
a net $S_{z}$ of $+(-)\frac{1}{2}$ since the wavefunction consists 
of $S_{tot}^{z}=0$ terms only. Under inversion of all spins in $e_{+}$ we get $e_{-}$. With this in mind, consider
the second sum on the right. Using the time reversal symmetry of the
wavefunction i.e. $w_{e_{+}}^{\downarrow\uparrow\downarrow}=zw_{e-}^{\uparrow\downarrow\uparrow}$
and $w_{e_{+}}^{\uparrow\downarrow\downarrow}=zw_{e-}^{\downarrow\uparrow\uparrow}$ (as seen from equation~\eqref{eq:w_relation}),
in equation~\eqref{eq:psi_S_psi} we get, 
\begin{eqnarray}
\left\langle \psi\right|S_{j}^{-}S_{k}^{+}S_{l}^{z}\left|\psi\right\rangle  & = & \frac{1}{2}\sum_{e_{-}}w_{e_{-}}^{\downarrow\uparrow\uparrow}w_{e_{-}}^{\uparrow\downarrow\uparrow}\\\nonumber&-&\frac{1}{2}z^{2}\sum_{e_{-}}w_{e_{-}}^{\downarrow\uparrow\uparrow}w_{e_{-}}^{\uparrow\downarrow\uparrow}\\
 & = & 0
\end{eqnarray}
where we have used $z^{2}=1$.

\renewcommand{\theequation}{{C}.\arabic{equation}}
\section{Schwinger Boson Mean Field Theory Calculations}
As mentioned in section~\ref{sec:SBMFT}, optimization of multiple parameters occurring in 
the Schwinger Boson theory for non uniform systems was quite a challenging task. 
Hence, for the interested reader, we report the exact values of the parameters 
obtained from our calculations, so that they may be able to reproduce our results.

The optimal mean field parameters are tabulated in Table~\ref{table:SBMFT_param} for different lattice sizes. 
In each column (from top to down) the parameters label inner to outermost most bonds/sites. 
The $Q_{ij}$'s alternate in strength across 
successive bonds consistent with the location of unpaired spins. 
Similar alternation in the condensate field $\beta_{i}$ indicates the variation 
in the density of dangling spins across shells. 

The ground state energy from SBMFT for the 126 site cluster was found to be $\approx -0.533 J$. This
is lower than the DMRG estimate $-0.39385 J$. This can be attributed to the well known fact~\cite{Auerbach,messio} 
about the non variational nature of SBMFT energies. This is because of not satisfying the constraints 
in equation~\eqref{eq:lambda constraint 0} exactly.

\begin{table}[h]
\begin{center}
\begin{tabular}{|c|c|c|c|c|c|c|c|}
\hline 
$\;\;\;\;N_{s}\;\;\;$ & $Q_{ij}^{*}$  & $\lambda_{i}^{*}$ & $\beta_{i}^{*}$ & $\;\;\;N_{s}\;\;\;$ & $Q_{ij}^{*}$ & $\lambda_{i}^{*}$ & $\beta_{i}^{*}$\tabularnewline
\hline 
 & 0.672 & 1.639 & 0 &  & $0.568$ & $1.893$ & $0.43$\tabularnewline
\cline{2-4} \cline{6-8} 
$ 14$ & 0.539 & 2.318 & 0 &  & $0.622$ & $1.656$ & $0.445$\tabularnewline
\cline{2-4} \cline{6-8} 
 & 0.676 & 0.523 & 0 & $ 126$ & $0.579$ & $2.026$ & $ $$0.392$\tabularnewline
\cline{1-4} \cline{6-8} 
 & 0.561 & 1.921 & 0 &  & $0.622$ & $1.497$ & $0.42$\tabularnewline
\cline{2-4} \cline{6-8} 
$ 30$ & 0.633 & 1.487 & 0 &  & $0.55$ & $2.32$ & $0.353$\tabularnewline
\cline{2-4} \cline{6-8} 
 & 0.543 & 2.345 & 0 &  & $0.677$ & $0.546$ & $0.443$\tabularnewline
\cline{2-8} 
 & 0.673 & 0.536 & 0 &  & $0.631$ & $1.669$ & $0.495$\tabularnewline
\cline{1-4} \cline{6-8} 
 & 0.646 & 1.72 & 0 &  & $0.56$ & $1.986$ & $0.454$\tabularnewline
\cline{2-4} \cline{6-8} 
 & 0.57 & 1.975 & 0 &  & $0.622$ & $1.537$ & $0.504$\tabularnewline
\cline{2-4} \cline{6-8} 
$ 62 $ & 0.63 & 1.514 & 0 & $ 254 $ & $0.571$ & $1.965$ & $0.455$\tabularnewline
\cline{2-4} \cline{6-8} 
 & 0.551 & 2.345 & 0 &  & $0.622$ & $1.495$ & $0.497$\tabularnewline
\cline{2-4} \cline{6-8} 
 & 0.68 & 0.544 & 0 &  & $0.55$ & $2.263$ & $0.449$\tabularnewline
\cline{1-4} \cline{6-8} 
\multicolumn{1}{|c}{} & \multicolumn{1}{c}{} & \multicolumn{1}{c}{} &  &  & $0.671$ & $0.563$ & $0.56$\tabularnewline
\hline 
\end{tabular}
\caption{Optimal SBMFT parameters for bond centered clusters of various sizes} 
\label{table:SBMFT_param}
\end{center}
\end{table}


\begin{thebibliography}{References}
\bibitem{FN_SMA} The expression $\sum_{k,l}u_{k}u_{l}G_{kl}$ is invariant under the transformation $u_{i}\rightarrow u_{i}+u$ though 
it is not manifestly evident at first glance. 
One can prove this, by using two results specific to a singlet state. 
First, $\sum_{ij}G_{ij}=S(S+1)$ which equals zero for a singlet state.
Second, $\sum_{ij}u_{j}G_{ij}=\sum_{j}\left\langle S^{tot}_{z} S^{z}_{j} \right \rangle$ which equals zero for a state with $S_{z}^{tot}=0$.
\end{thebibliography}
\end{document}